\pdfoutput=1
\RequirePackage{ifpdf}
\ifpdf 
\documentclass[pdftex]{sigma}
\else
\documentclass{sigma}
\fi

\numberwithin{equation}{section}

\newtheorem{Theorem}{Theorem}[section]
\newtheorem{Corollary}[Theorem]{Corollary}
\newtheorem{Lemma}[Theorem]{Lemma}
\newtheorem{Proposition}[Theorem]{Proposition}
 { \theoremstyle{definition}
\newtheorem{Example}[Theorem]{Example}
\newtheorem{Remark}[Theorem]{Remark} }

\let\trueint=\int
\def\int{\mathop{\textstyle\trueint}\limits}
\let\<=\langle
\let\>=\rangle

\def\RR{{\mathcal R}}

\def\Sp{{\rm Spec}}

\def\0{\mathbf{0}}

\def\bb{\hskip0.3cm\circle*{6}}

\usepackage{epstopdf}
\usepackage{tikz}
\usepackage{amsbsy,amsfonts,amscd}
\usepackage[mathscr]{eucal}
\usepackage{young}

\newcommand{\Z}{\mathbb Z}

\newcommand{\N}{\mathbb N}
\newcommand{\C}{\mathbb C}

\begin{document}

\newcommand{\arXivNumber}{1912.06768}

\renewcommand{\PaperNumber}{024}

\FirstPageHeading

\ShortArticleName{Space Curves and Solitons of the KP Hierarchy.~I}

\ArticleName{Space Curves and Solitons of the KP Hierarchy.\\ I.~The $\boldsymbol{l}$-th Generalized KdV Hierarchy}

\Author{Yuji KODAMA and Yuancheng XIE}

\AuthorNameForHeading{Y.~Kodama and Y.~Xie}

\Address{Department of Mathematics, The Ohio State University, Columbus OH, 43210, USA}
\Email{\href{mailto:kodama.1@osu.edu}{kodama.1@osu.edu}, \href{mailto:xie.635@osu.edu}{xie.635@osu.edu}}

\ArticleDates{Received October 14, 2020, in final form March 02, 2021; Published online March 16, 2021}

\Abstract{It is well known that algebro-geometric solutions of the KdV hierarchy are constructed from the Riemann theta functions associated with hyperelliptic curves, and that soliton solutions can be obtained by rational (singular) limits of the corresponding curves. In this paper, we discuss a class of KP solitons in connections with space curves, which are labeled by certain types of numerical semigroups. In particular, we show that some class of the (singular and complex) KP solitons of the $l$-th generalized KdV hierarchy with $l\ge 2$ is related to the rational space curves associated with the numerical semigroup $\langle l,lm+1,\dots, lm+k\rangle$, where $m\ge 1$ and $1\le k\le l-1$. We~also calculate the Schur polynomial expansions of the $\tau$-functions for those KP solitons. Moreover, we construct smooth curves by deforming the singular curves associated with the soliton solutions. For~these KP solitons, we also construct the space curve from a commutative ring of differential operators in the sense of the well-known Burchnall--Chaundy theory.}


\Classification{37K40; 37K10; 14H70; 14H50}

\section{Introduction}

The purpose of this paper is to identify algebraic curves associated with particular class of soliton solutions of the KP hierarchy. It is known that solutions of the KP equation can be constructed from \emph{any} algebraic curves~\cite{Kr:77}. A solution from a smooth curve is a \emph{quasi-periodic} solution, and some soliton solutions can be constructed by a rational (singular) limit of the curve with only ordinary double points (see, e.g.,~\cite{BBEIM:94, Mu:84, SW:85}). In particular, the cases corresponding to the KdV and nonlinear Sch\"odinger equations are well-studied, in which the algebraic curves are given by the hyperelliptic curves (see, e.g.,~\cite{BBEIM:94, Mu:84}). Recently, there are several papers dealing with some
non-hyperelliptic cases, e.g., so-called $(n,s)$-curves, where the authors construct the Klein $\sigma$-functions over these curves (see, e.g.,~\cite{BLE:97, BLE:99, KMP:13, KMP:16, KMP:18, MK:13, Na:10}).
It seems, however, that almost no result has been reported for the cases with more general algebraic curves. Because of the difficulty in finding a canonical homological basis for the general algebraic curves, it may be quite complicated to compute
explicitly a rational limit of these curves and the cor\-res\-pon\-ding Riemann theta functions (see~\cite{Na:10}). On the other hand, a large number of real regular soliton solu\-ti\-ons of~the KP hierarchy has been classified in terms of finite Grassmann manifolds ${\rm Gr}(N,M)$ (see, e.g.,~\cite{K:17, KW:13, KW:14}). Recently, there are some progress on the study concerned with the connections between the algebraic-geometric solutions and these soliton solutions~\cite{Ab:17, AG:18, Na:18a, Na:18b}.
 	
In this paper, we take the first step to identify explicit forms of algebraic curves associated with certain class of KP solitons. We~consider complex (singular in general) soliton solutions
	in~terms of the complex Grassmannians. Here what we mean by ``soliton'' is that the underlying
	$M$-dimensional space $\C^M$ of the Grassmannian for the solution expressed by the so-called
	\emph{$\tau$-function} is spanned by only ``exponential'' functions. We~also introduce the ``generalized" soliton solutions which are mixed with rational functions in addition to the exponential functions (see~\cite{Na:18a}), which are obtained by multiple degenerations of the soliton parameters.
	In particular, choosing a special point of the Schubert cell of the Grassmannian, the $\tau$-function corresponding to a generalized soliton has a Schur polynomial expansion whose leading polynomial is determined by the Young diagram corresponding to the Schubert cell.
	The main tool in this paper is the Sato universal Grassmannian which is the moduli space of the formal power series solutions of the KP hierarchy~\cite{Na:18a, Na:18b, S:81, SN:84}.
	
\looseness=1 We briefly give a background of the Sato theory of the KP hierarchy and the algebro-geometric solutions associated with algebraic curves. Sato's theory of KP hierarchy associates a point of~the
universal Grassmannian manifold (UGM) with a solution of the KP hierarchy. Each solution of
the KP hierarchy is given by the so-called $\tau$-function, which is a function of infinitely many variables $(t_n\colon n=1,2,\ldots)$. It is known that an algebro-geometric solution of the KP hierarchy can be expressed by the Riemann theta function~\cite{BBEIM:94, Kr:77, Ml:84}. A method of algebro-geometric solutions was proposed in~\cite{Kr:77}, and
these solutions are constructed by the following way.
Let~$\mathcal{C}$ be a compact Riemann surface of genus $g$ and $\mathcal{L}$ a holomorphic line bundle of degree $g$ over~it. Let~$p_{\infty}$ be a nonsingular point of $\mathcal{C}$, $z$ a local parameter around $p_{\infty}$. Let~$H^0(\mathcal{C}, \mathcal{L}(*\infty))$ be meromorphic sections of $\mathcal{L}$ with poles only at $p_{\infty}$. Identifying $\mathcal{L}$ with a degree $g$ divisor $D$ on~$\mathcal{C}$, this can be represented as a space of meromorphic functions $U$ on a small neighborhood of $p_{\infty}$.
Such $U$ represents a point of UGM, and denoted by $U={\rm Span}_\C\{\mathcal{B}\}$ with the Sato frame $\mathcal{B}$ representing a basis of the space of meromorphic functions
with poles at the specified divisors $D+*\infty$ on $\mathcal{C}$. Then the Sato frame representing a point of UGM determines the coefficients of~the Schur polynomial expansion of the $\tau$-function.
In the present paper, we only consider the set of meromorphic functions with a pole only at $p_\infty$, i.e., $D=gp_\infty$. (The general case will be discussed in a future communication~\cite{YX:21})
Let~$W = H^{0}(\mathcal{C}, \mathcal{O}(*\infty))$ be the set of meromorphic functions on $\mathcal{C}$ with a pole only at $p_{\infty}$. We~note that $W$ is a commutative algebra graded by the order of poles at $p_{\infty}$ and $U$ is a module over $W$, where we have
$z^gW=U\in\text{UGM}$~\cite{SW:85}. We~also note that the order of poles of elements in $W$ defines a numerical semi\-group~$S$.
	
It turns out that the algebraic curve associated with the generalized soliton solution discussed in this paper is a ``singular space curve''. The strategy to find the curve is as follows.
\begin{enumerate}\itemsep=0pt
\item 	Given a $\tau$-function associated with a generalized soliton solution, we identify the cor\-res\-ponding Sato frame $\mathcal{B}$ by embedding the solution into UGM as a point $U={\rm Span}_\C\{\mathcal{B}\}$.
\item We define a basis $\mathcal{A}$ for a maximal graded commutative subalgebra $\RR$
in $\C\big[z^{-1}\big]$, over which $U$ is an $\RR$-module. The basis $\mathcal{A}$ is filtered by the order of poles at $p_{\infty}$,
and in the cases under consideration it has the form $\mathcal{A}=z^{-g}f(z)\mathcal{B}$ with some polynomial $f(z)$ with $f(0)=1$.
\item We identify a (singular) space curve $\mathcal{C}$ whose affine part is given by
$\text{Spec}(\RR)$. In the case of a smooth curve $\tilde{\mathcal{C}}$, there is an identification,
${\rm Span}_\C\{\mathcal{A}\}=H^0\big(\tilde{\mathcal{C}},\mathcal{O}(*\infty)\big)$ and $z^gH^0\big(\tilde{\mathcal{C}}, \mathcal{O}(*\infty)\big)\in\text{UGM}$.
The numerical semigroup $S$ represents the order of poles of the basis $\mathcal{A}$, and the genus of $S$ is the arithmetic genus of the curve~$\mathcal{C}$.
\end{enumerate}
	
The present paper is organized as follows. In Section~\ref{sec:flag}, we define a class of generalized soliton solutions of the KP hierarchy which are defined by further degenerations of the soliton parameters. These solitons are particular solutions of the $l$-th generalized KdV hierarchy for some integer $l\ge 2$, which
is also referred to as the $l$-reduction of the KP hierarchy defined by $\partial\tau/\partial t_{nl}=c_n\tau$ with some constants $c_n$ for all $n=1,2,\ldots$. In particular, we consider the generalized soliton solutions
from the \emph{center} elements of the Schubert cell of the Grassmannian ${\rm Gr}(N,M)$
(Theorem~\ref{thm:center}).

In Section~\ref{sec:SG}, we briefly review the notion of Sato universal Grassmannian (UGM) and give an explicit embedding of the generalized soliton solutions into UGM. Each soliton solution is then expressed as a point of UGM (Proposition~\ref{prop:phi-f}).

In Section~\ref{sec:NS}, we study the numerical semigroups of type
$\langle l,lm+1,lm+2,\dots,lm+k\rangle$ for~$m\ge 1$ and $1\le k\le l-1$, and present the corresponding Young diagrams, which will be used to parametrize the generalized soliton solutions
to the $l$-th generalized KdV hierarchy. We~also give the explicit formulas of the Frobenius number and (arithmetic) genus of these numerical semigroups (Proposition~\ref{prop:FG}).

In Section~\ref{sec:SGlmk}, we give the main theorem (Theorem~\ref{thm:Main}), which states that algebraic curves for~the generalized soliton solutions are
singular space curve in $\C^{k+1}$ associated with the numerical semigroup of type $\langle l,lm+1,\dots,lm+k\rangle$. Theorem~\ref{thm:Main} also provides the Schur expansions of the corresponding $\tau$-function. We~then present several explicit examples to illustrate the theorem.

In Section~\ref{sec:Proof}, we give the proof of Theorem~\ref{thm:Main}. This section also provides a summary of our main results.

In Section~\ref{sec:deformation}, we study a deformation
of the singular space curve in $\C^l$ for the soliton solution associated with the numerical semigroup $S=\langle l,lm+1,\ldots,l(m+1)-1\rangle$ (Proposition~\ref{prop:smooth}). It is then shown that these soliton solutions are obtained by coalescing two moduli parameters of the corresponding smooth curve. The singularity of the singular space curve is an~ordinary $l$-tuple point singularity (see, e.g.,~\cite{BG:80}),
and in the case with $l=2$, i.e., a plane curve, it is an~ordinary ``double point'' singularity. As a byproduct, we show that the numerical semigroup of type $\langle l,lm+1,\ldots,l(m+1)-1\rangle$ is a Weierstrass semigroup for $m\ge 2$. We~also give a brief discussion on deformation of singular curves for the generalized soliton solutions
associated with the types $\langle l, lm+1,\ldots,lm+k\rangle$ with $1\le k<l-1$.

Finally, in Section~\ref{sec:spectralC}, we reconstruct the singular space curves for the soliton solution
 from the point of view of commutative differential
operators as in the work of Burchnall and Chaundy \cite[Theorem~\ref{thm:SpecEV}]{BC:23}.

In the appendix, we provide a brief background information about the Lax--Sato formulation of the KP hierarchy, the $l$-th generalized KdV hierarchy and the soliton solutions.

\section[The generalized soliton solutions and the Schur expansion of~the tau-function]{The generalized soliton solutions\\ and the Schur expansion of~the $\boldsymbol \tau$-function}\label{sec:flag}

The KP equation is a nonlinear partial differential equation in the form,
\begin{gather}\label{eq:KP}
\partial_1\big({-}4\partial_3 u+6u\partial_1u+\partial_1^3u\big)+3\partial_2^2u=0,
\end{gather}
with the partial derivatives $\partial_n^ku=\frac{\partial^ku}{\partial t_n^k}$.
The KP equation admits an infinite number of symmetries (commuting flows), and
these flows are parametrized by the ``time'' variables $t_n$ for~$n\in \N$.
The set of all commuting flows is called the KP hierarchy, whose first member is the KP equation.

The KP soliton solutions are classified in terms of finite dimensional Grassmannians.
Here we briefly summarize these previous results for the background of the present paper
(also see Appendix~\ref{A}).

\subsection[The tau-functions for the generalized soliton solutions]{The $\boldsymbol\tau$-functions for the generalized soliton solutions}

The $\tau$-function of the KP hierarchy with the time variables $t=(t_1,t_2,\ldots)$ is introduced by
\begin{gather}\label{eq:u}
u(t)=2\partial_1^2\ln \tau(t).
\end{gather}
The soliton solutions are constructed as follows.
Let~$\{f_i(t)\colon 1\le i\le N\}$ be a set of linearly independent functions $f_i(t)$ of
infinite variables $t=\{t_n\colon n\in \N\}$ satisfying the following system of linear equations,
\begin{gather}\label{eq:f}
\partial_nf_i=\partial_1^nf_i\qquad\text{for}\quad 1\le i\le N\quad \text{and}\quad n\in\N.
\end{gather}
The $\tau$-function is then given by the Wronskian with respect to $t_1$-variable,
\begin{gather}\label{eq:Wr}
\tau(t)=\text{Wr}(f_1,f_2,\ldots,f_N).
\end{gather}
(See~\eqref{eq:WrA} in Appendix~\ref{A}, and also~\cite{K:17} for the details.)

As a fundamental set of the solutions of~\eqref{eq:f}, we take the exponential functions,
\begin{gather*}\label{eq:E}
E_j(t)={\rm e}^{\theta_j(t)}\qquad
\text{with}\quad
\theta_j(t):=\sum_{n=1}^\infty\kappa_j^nt_n\qquad
\text{for}\quad j=1,\ldots,M,
\end{gather*}
where $\kappa_j$'s are arbitrary constants.
For the soliton solutions, we consider $f_i(t)$ as a linear combination of the exponential solutions,
\begin{gather}\label{eq:Ef}
f_i(t)=\sum_{j=1}^Ma_{i,j}E_j(t)\qquad\text{for}\quad i=1,\ldots,N,
\end{gather}
where $A:=(a_{i,j})$ is an $N\times M$ constant matrix of full rank, ${\rm rank}(A)=N$,
Then the $\tau$-fun\-c\-tion~\eqref{eq:Wr} is expressed by
\begin{gather}\label{eq:tauE}
\tau(t)=\big|E(t) A^{\rm T}\big|,
\end{gather}
where $A^{\rm T}$ is the transpose of the matrix~$A$, and $E(t)$ is given by
\begin{gather}\label{eq:Et}
E(t)=\begin{pmatrix}
E_1 & E_2 &\cdots & E_M\\
\partial_1E_1&\partial_1E_2&\cdots &\partial_1E_M\\
\vdots &\vdots &\ddots &\vdots\\
\partial_1^{N-1}E_1 &\partial_1^{N-1}E_2&\cdots &\partial_1^{N-1}E_M
\end{pmatrix}\!.
\end{gather}
Note here that the set of exponential functions $\{E_1(t),\ldots,E_M(t)\}$ gives a basis of the $M$-di\-men\-sional space
of the null space of the operator $\prod_{i=1}^M(\partial_1-\kappa_i)$, and we call it a ``basis''
of~the~KP soliton. Then the set of functions $\{f_1(t),\ldots,f_N(t)\}$ represents an $N$-dimensional subspace
of $M$-dimensional space spanned by the exponential functions. This leads naturally to~the structure
of a finite Grassmannian ${\rm Gr}(N,M)$, the set of $N$-dimensional subspaces in $\C^M$.
Then the $N\times M$ matrix~$A$ of full rank can be identified as a point of ${\rm Gr}(N,M)$.
The ${\rm Gr}(N,M)$ has a decomposition called the Schubert decomposition given by
\begin{gather}\label{eq:Schubert}
{\rm Gr}(N,M)=\bigsqcup_{\lambda\subset (M-N)^N}X_\lambda,
\end{gather}
where $\lambda$ is a Young diagram contained in an $N\times (M-N)$ rectangular diagram
denoted by $(M-N)^N$, and $X_\lambda$ is the Schubert cell which consists of the $N\times M$
matrices whose pivots are given by $1\le i_1<i_2<\cdots <i_N\le M$ with $i_k=M-N+k-\lambda_k$.
In terms of the pivot indices, the Young diagram $\lambda=(\lambda_1,\ldots,\lambda_N)$
is expressed as
\vspace{1ex}
\setlength{\unitlength}{0.5mm}
\begin{center}
 \begin{picture}(120,60)
\put(5,55){\line(1,0){142}}
\put(5,45){\line(1,0){110}}
 \put(5,35){\line(1,0){90}}
 \put(5,25){\line(1,0){90}}
 \put(5,15){\line(1,0){70}}
 \put(5,5){\line(1,0){70}}
 \put(5,5){\line(0,1){50}}
 \put(31,5){\line(0,1){50}}
 \put(18,5){\line(0,1){50}}
 \put(75,5){\line(0,1){10}}
 \put(62,5){\line(0,1){10}}
 \put(95,25){\line(0,1){10}}
 \put(82,25){\line(0,1){10}}
\put(115,45){\line(0,1){10}}
\put(102,45){\line(0,1){10}}
 \put(98,29){${i_{n}}$}
 \put(-35,29){$M-N+n$}
 \put(-35,49){$M-N+1$}
 \put(-13,38){$\vdots$}
 \put(-13,18){$\vdots$}
 \put(65,38){$\vdots$}
 \put(10,38){$\vdots$}
 \put(24,38){$\vdots$}
 \put(100,40){${i_1+1}$}
 \put(60,49){$\cdots$}
 \put(60,58){$\cdots$}
 \put(55,29){$\cdots$}
 \put(47,18){$\vdots$}
 \put(10,18){$\vdots$}
 \put(24,18){$\vdots$}
 \put(80,20){${i_n+1}$}
 \put(-16,9){$M$}
 \put(45,9){$\cdots$}
 \put(78,9){${i_N}$}
 \put(2,58){$M-N$}
 \put(105,58){${i_1}{\qquad\cdots\qquad 1}$}
 \put(118,49){${i_1}{\quad\cdots\quad ~1}$}

 \put(7,-1){$ M$}
 \put(18,-1){${M-1\quad\cdots}$}
 \put(60,-1){${i_N+1}$}
 \end{picture}
\end{center}

\noindent
That is, the pivot indices of the matrices in $X_\lambda$ appear at the vertical paths in a lattice path starting from the top right corner and ending at the bottom left corner in the Young diagram $\lambda$ as shown in the figure above.

We also have the following formula of the $\tau$-function using the Binet--Cauchy lemma (see, e.g.,~\cite{K:17}),
\begin{gather*}
\tau(t)=\sum_{I\in\binom{[M]}{N}}\Delta_{I}(A)E_{I}(t),
\end{gather*}
where $I=\{i_1<i_2<\cdots<i_N\}$ is an $N$ element subset in $[M]=\{1,2,\ldots,M\}$, $\Delta_{I}(A)$ is the $N\times N$ minor with the column vectors indexed by $I=\{i_1,\ldots,i_N\}$,
and $E_{I}(t)$ is the $N\times N$ determinant of the same set of the columns in~\eqref{eq:Et}, which is given by
\begin{gather*}\label{eq:EI}
E_{I}(t)=\prod_{k<l}(\kappa_{i_l}-\kappa_{i_k})\,E_{i_1}(t)\cdots E_{i_N}(t).
\end{gather*}
The minor $\Delta_{I}(A)$ is also called the Pl\"ucker coordinate, and the $\tau$-function represents a
point of~${\rm Gr}(N,M)$ in the sense of the Pl\"ucker embedding, ${\rm Gr}(N,M)\hookrightarrow \mathbb{P}\big({\wedge}^N\C^M\big)\colon A\mapsto \big\{\Delta_I(A)\colon$ $I\in\binom{[M]}{N}\big\}$
(see, e.g.,~\cite{K:17}).

\subsection[The Schur function expansion of the tau-function]{The Schur function expansion of the $\boldsymbol\tau$-function}

Expanding the exponential function $E_j(t)$ with respect to the powers of $\kappa_j$, we have
\begin{gather*}
E_j(t)={\rm e}^{\theta_j(t)}=\sum_{n=0}^\infty \kappa_j^np_n(t),
\end{gather*}
where $p_n(t)$ is the elementary Schur polynomials. That is, $E_j(t)$ is the generating function of
the elementary Schur polynomials, which are given by
\begin{gather*}
p_n(t)=\sum_{i_1+2i_2+\cdots+ni_n=n}\frac{t_1^{i_1}t_2^{i_2}\cdots t_n^{i_n}}{i_1!i_2!\cdots i_n!}.
\end{gather*}
Note that the elementary Schur polynomials satisfy
\begin{gather*}
\partial_kp_n(t)=\partial_1^kp_n(t)=p_{n-k}(t)\qquad
\text{and}\qquad
p_m(t)=0\qquad\text{if}\quad m<0.
\end{gather*}

Using the elementary Schur polynomials, the $\tau$-function can be expressed as follows.
First note that
\begin{gather*}
(f_1,f_2,\ldots, f_N)=(E_1,E_2,\ldots, E_M)A^{\rm T},
\end{gather*}
where $A^{\rm T}$ is the transpose of the $N\times M$ matrix~$A$. Expanding the exponential functions, we~have
\begin{gather*}
(E_1,E_2,\ldots,E_M)=(1,p_1,p_2,\ldots)K,
\end{gather*}
where $K$ is the $\infty\times M$ Vandermonde matrix, $K=\big(\kappa_j^{i-1}\big)$ for
$1\le i$ and $1\le j\le M$, i.e.,
\begin{gather}\label{eq:K}
K=\begin{pmatrix}
1 & 1& \cdots & 1\\
\kappa_1&\kappa_2&\cdots&\kappa_M\\
\kappa_1^2&\kappa_2^2&\cdots&\kappa_M^2\\
\vdots &\vdots& \vdots &\vdots
\end{pmatrix}\!.
\end{gather}
Note that each function $f_i(t)$ in~\eqref{eq:Ef} has the following expansion,
\begin{gather*}\label{eq:fi}
f_i(t)=\sum_{n=0}^\infty\sum_{j=1}^M a_{i,j}\kappa_j^np_n(t)\qquad\text{for}\quad i=1,2,\ldots,N.
\end{gather*}
The $\tau$-function~\eqref{eq:tauE} is then expressed by
\begin{gather}\label{eq:tauP}
\tau(t)=|EA^{\rm T}|=\left|
\begin{pmatrix}
1& p_1&p_2&\cdots&\cdots &\cdots&\cdots\\
0& 1 & p_1 &p_2&\cdots&\cdots&\cdots\\
\vdots&\ddots&\ddots &\ddots &\ddots&\ddots&\vdots\\
0&\cdots& 0&1&p_1&p_2 & \cdots
\end{pmatrix}\, K\, A^{\rm T}\right|.
\end{gather}

In this paper, we include the \emph{generalized} bases which
contain derivatives of the exponential functions with respect to the parameters $\kappa_j$
\cite{Na:18a}. For example, we consider the following derivative,
\begin{gather}\label{eq:gE}
\hat E_j^{(q)}(t):=\frac{1}{q!}\frac{\partial^q}{\partial\kappa_j^q}\big(\kappa_j^qE_j(t)\big)=\sum_{n=0}^\infty
\binom{q+n}{n}\kappa_j^np_n(t).
\end{gather}
Note that $\hat E_j^{(0)}=E_j$.
Then one can take the following basis,
\begin{gather}\label{eq:gEset}
\big(\hat E_1^{(q_1)}(t),\hat E_2^{(q_2)}(t),\ldots,\hat E_M^{(q_M)}(t)\big),
\end{gather}
where $q_j$'s are nonnegative integers. The $\tau$-function in this case can be expressed in
the form~\eqref{eq:tauP} by replacing the Vandermonde matrix $K$ in~\eqref{eq:K} with
\begin{gather}\label{eq:Kg}
\hat K=\begin{pmatrix}
1 & 1 & \cdots & 1\\[0.5ex]
\binom{q_1+1}{1}\kappa_1 &\binom{q_2+1}{1}\kappa_2 &\cdots &\binom{q_M+1}{1}\kappa_M\\[1.5ex]
\binom{q_1+2}{2}\kappa_1^2 &\binom{q_2+2}{2}\kappa_2^2 &\cdots &\binom{q_M+2}{2}\kappa_M^2\\
\vdots &\vdots &\vdots &\vdots
\end{pmatrix}
\end{gather}
which we call the \emph{generalized} Vandermonde matrix. Then the functions $\hat f_i(t)$ for the $\tau$-function, denoted by $\hat\tau=\text{Wr}(\hat f_1,\ldots,\hat f_N)$, have the expansion,
\begin{gather}\label{eq:gfi}
\hat f_i(t)=\sum_{j=1}^Ma_{i,j}\hat E_j^{(q_j)}(t)=\sum_{n=0}^\infty\sum_{j=1}^Ma_{i,j}\binom{q_j+n}{n}\kappa_j^np_n(t).
\end{gather}

\begin{Remark}
The soliton solutions generated by the $\tau$-function with the basis~\eqref{eq:gE} contain
rational functions of the variables $t=(t_1,t_2,\ldots)$, hence they are singular in general.
\end{Remark}

Now we consider the expansion of the $\tau$-function, which we call the Schur expansion of the $\tau$-function:
Using the Binet--Cauchy lemma, one can expand the $\tau$-function~\eqref{eq:Wr} in terms of
the Schur polynomials,
\begin{gather}\label{eq:SchurExp}
\tau(t)=S_\lambda(t)+\sum_{\hat\mu\supset\lambda}c_{\hat\mu} S_\mu(t),
\end{gather}
where $S_\lambda(t)$ is the Schur polynomial associated with the Young diagram $\lambda=(\lambda_1,\lambda_2,\ldots,\lambda_N)$ defined by
\begin{gather*}
S_\lambda(t)=|(p_{\lambda_i+j-i}(t))|.
\end{gather*}
The constant $c_{\hat\mu}$ in~\eqref{eq:SchurExp} are the $N\times N$ minors of the $\infty\times N$ matrix $KA^{\rm T}$ (or $\hat KA^{\rm T}$).
In~particular, the first constant $c_\lambda$, which is normalized to be one, is the minimal minor
of $KA^{\rm T}$ (or~$\hat KA^{\rm T}$) in the lexicographical order of the row indices $\{l_0,l_1,\ldots,l_{N-1}\}$ for $0\le l_0<l_1<\cdots<l_{N-1}$
with the relation,
\begin{gather*}
l_k=\lambda_{N-k}+k\qquad\text{for}\quad 0\le k\le N-1.
\end{gather*}
With this relation, the Schur polynomial associated with the Young diagram $\lambda$ is given by
\begin{gather}\label{eq:SchurP}
S_\lambda(t)=\text{Wr}(p_{\lambda_N},p_{\lambda_{N-1}+1},\ldots,p_{\lambda_1+N-1})
=\text{Wr}(p_{l_0},p_{l_1},\ldots,p_{l_{N-1}}).
\end{gather}

Notice that in~\eqref{eq:SchurExp} if $A$ is totally nonnegative, then $\lambda=\varnothing$ and
$\tau(0)=1$ (note that a~real KP soliton is regular, if and only if $A$ is totally nonnegative~\cite{KW:13}). Here, we are interested in~the singular solutions, where $\tau(0)=0$. In particular, we look for the expansion~\eqref{eq:SchurExp} with a~particular matrix~$A$ so that the leading Schur polynomial has the largest possible weight. Recall that the weight of $S_\lambda(t)$ is defined by the number of boxes in $\lambda$, denoted by $|\lambda|=\sum_{i=1}^N\lambda_i$. Note that the weight of $t_n$ is $n$.

We then have the following theorem, which identifies the leading Schur polynomial of the $\tau$-function for each Schubert cell associated with the Young diagram $\lambda$.
\begin{Theorem}\label{thm:center}
Fixed a Schubert cell $X_{\lambda}\subset {\rm Gr}(N,M)$, there exists a unique element $A\in X_\lambda$, so that the leading term of the Schur expansion of the $\tau$-function~\eqref{eq:tauP}
generated by $A$ and $K$ in~\eqref{eq:K} $($or $\hat K$ in~\eqref{eq:Kg}$)$ is $S_{\lambda}(t)$.
Here, the pivot indices, say $\{i_1<i_2<\cdots<i_N\}$, of the matrix~$A$ are related to $\lambda$ with
\begin{gather*}
i_k=M-N+k-\lambda_k\qquad\text{for}\quad k=1,\ldots,N.
\end{gather*}
\end{Theorem}
\begin{proof}
Assume $A$ is in an row echelon form, i.e.,
\begin{gather*}
A = \begin{pmatrix}
	\begin{array}{ccccccccccccccccc}
	0 & \cdots & 0 & \young[1][6][1] & * & \cdots & * & * & * & \cdots & \cdots & * & * & \cdots&* \\
	0 & \cdots & 0 & 0 & 0 & \cdots & 0 & \young[1][6][1] & * & \cdots & \cdots & * & * & \cdots&* \\
	\vdots & & \vdots & \vdots & \vdots& & \vdots & \vdots & \vdots & \vdots &\vdots & \vdots & \vdots &\vdots&\vdots \\
	0 & \cdots & 0 & 0 & 0 & \cdots & 0 & 0 & 0 & \cdots& \cdots & 0 & \young[1][6][1] & \cdots&*
	\end{array}
	\end{pmatrix}\!,
	\end{gather*}
where $\young[1][6][1]$ are the pivot one's. Note that the $k$-th row of $A$ has exactly $M - i_k = l_{N-k}$ many free parameters indicated by ``$*$" in $A$. Since the rows of the $K$-matrix in~\eqref{eq:K} (or $\hat K$ in~\eqref{eq:Kg}) are linearly independent, we can use these free parameters to annihinate the first $l_{N-k}$ entries in~the $k$-th column of $KA^{\rm T}$ in~\eqref{eq:tauP}, and such $A$ matrix is uniquely determined. We~can then use column operation of $KA^{\rm T}$ to make the first nonzero entries in each column to be $1$. Using the Binet--Cauchy formula, we have that the first nonzero term in the Schur expansion of $\tau$-function is given by~\eqref{eq:SchurP}.
\end{proof}

\begin{Remark}
One should note in general that for an $A\in X_\lambda$, the $\tau$-function has
a Schur expansion with the leading polynomial $S_\mu(t)$, whose Young diagram satisfies $\mu\subset \lambda$. In the case of $\lambda=\mu$ in Theorem~\ref{thm:center},
the unique matrix~$A$ (more precisely, the reduced row echelon form of $A\in X_\lambda$) is sometimes referred to as the ``center'' of the cell $X_\lambda$. Note that the center $A$ also depends
on $K$ (or $\hat K$).
\end{Remark}

\subsection[The l-th generalized KdV hierarchy and the generalized soliton solutions]{The $\boldsymbol l$-th generalized KdV hierarchy and the generalized soliton solutions}

We consider the solutions for the $l$-th generalized KdV hierarchy (sometimes referred to as the $l$-reductions of the KP hierarchy), which is defined by
the following additional conditions for the functions $\{f_i(t)\colon i=1,\ldots,N\}$ in~\eqref{eq:f},
\begin{gather}\label{eq:reduction}
\partial_{nl}f_i=r_i^nf_i\qquad\text{for}\quad n=1,2,\ldots,
\end{gather}
where $r_i$'s are some constants. We~consider the following set of $f_i$'s. Let~$\{n_1,\ldots,n_m\}$ be a~partition of $N$ with
$1\le n_s< l$, i.e., $N_s=n_1+\cdots+n_{s}$. Take
\begin{gather*}\label{eq:fil}
f_{i_s}(t)=\sum_{j=1}^la_{i_s,j}E_j(t,\kappa_s)\qquad \text{for}\quad
N_{s-1}+1\le i_s\le N_s \qquad
(1\le s\le m)
\end{gather*}
with $N_0=0$, $N_m=N$ and
\begin{gather}\label{eq:Ej}
 E_j(t,\kappa_s)=\exp\bigg(\sum_{n=1}^\infty \big(\kappa_s\omega_l^{j-1}\big)^nt_n\bigg)\qquad\text{for}\quad \begin{cases}
 1\le j\le l,\\[0.5ex]
 1\le s\le m.
 \end{cases}
\end{gather}
Here we have $r_{i_s}=\kappa_s^l$ and $\omega_l=\exp(2\pi {\rm i}/l)$, the $l$-th root of unity.
Then the $N\times ml$ matrix~$A$ consists of $m$ diagonal blocks in the form,
\begin{gather}\label{eq:A1}
A=\begin{pmatrix}
A[n_1] & 0 & \cdots & 0\\
0 & A[n_2] &\cdots &0\\
\vdots&\vdots&\ddots &\vdots\\
0&0&\cdots & A[n_m]
\end{pmatrix}\!.
\end{gather}
Each block $A[n_s]$ is an $n_s\times l$ matrix ($n_s<l$),
\begin{gather*}
A[n_s]=\begin{pmatrix}
a_{N_{s-1}+1,1}& a_{N_{s-1}+1,2} &\cdots & a_{N_{s-1}+1,l}\\
a_{N_{s-1}+2,1}&a_{N_{s-1}+2,2}&\cdots &a_{N_{s-1}+2,l}\\
\vdots &\vdots&\ddots&\vdots\\
a_{N_s,1}&a_{N_s,2}&\cdots &a_{N_s,l}
\end{pmatrix}\qquad \text{for}\quad 1\le s\le m.
\end{gather*}
 Note that the matrix~$A$ corresponds to the basis of functions,
\begin{gather*}
\big(E_1(t,\kappa_1),\ldots,E_l(t,\kappa_1),E_1(t,\kappa_2),\ldots,E_l(t,\kappa_2),\ldots,
E_1(t,\kappa_m),\ldots,E_l(t,\kappa_m)\big),
\end{gather*}
which we simply write as
\begin{gather}\label{eq:baseF}
\big(\mathsf{E}(\kappa_1),\mathsf{E}(\kappa_2),\ldots,\mathsf{E}(\kappa_m)\big),\qquad \text{where}\quad \mathsf{E}(\kappa_j)=(E_1(t,\kappa_j),\ldots,E_l(t,\kappa_j)).
\end{gather}

With the reduction~\eqref{eq:reduction}, we have
\begin{gather*}
f_{i_s}(t)=g_{i_s}(\hat t)\exp\bigg(\sum_{n=1}^\infty \kappa_s^{nl}t_{nl}\bigg)\qquad\text{for}\quad s=1,\ldots,m,
\end{gather*}
where $\hat t$ is the set of time variables without $t_{nl}$ for $n=1,2,\ldots$. Then the $\tau$-function
\eqref{eq:Wr} is expressed by
\begin{gather*}
\tau(t)=\exp\bigg(\sum_{s=1}^m\sum_{n=1}^\infty \kappa_s^{nl}t_{nl}\bigg)\text{Wr}(g_1,\ldots,g_N)(\hat t).
\end{gather*}
Since the solution of the KP hierarchy is given by~\eqref{eq:u}, the exponential factor of
the $\tau$-function does not give any contribution to the solution $u$. Then the $\tau$-function,
denoted by $\tau_l(\hat t)$,
for the $l$-reduced KP hierarchy can be expressed by
\begin{gather*}
\tau_l(\hat t)=\text{Wr}(g_1,\ldots,g_N)(\hat t)=\tau(t)\big|_{t_{nl}=0}.
\end{gather*}


\subsubsection[The generalized soliton solutions for the l-th generalized KdV hierarchy]
{The generalized soliton solutions for the $\boldsymbol l$-th generalized KdV hierarchy}

One should first note that the
reduction condition~\eqref{eq:reduction} is not compatible with the derivatives~\eqref{eq:gE}, that is,
the derivative $\hat E_j^{(q)}(t,\kappa)=\frac{1}{q!}\frac{\partial^q}{\partial \kappa^q}(\kappa^qE_j(t,\kappa))$ does not satisfy the condition~\eqref{eq:reduction} for $q>0$.
In order to construct the generalized soliton solutions for the reduced equation,
we consider a limit $\kappa_i\to\kappa_j$ for fixed $\kappa_j$.
For the sake of simplicity, we take $\kappa_1=\kappa_2+\Delta\kappa$.
Then the elements in the first $l$ terms in the set~\eqref{eq:baseF}, $\{E_1(t,\kappa_1),\ldots, E_l(t,\kappa_1)\}$, have the expan\-sions,
\begin{gather*}
E_k(t,\kappa_1)= E_k(t,\kappa_2)+\Delta \kappa\frac{\partial }{\partial \kappa_2}E_k(t,\kappa_2)+O\big(\Delta\kappa^2\big)\qquad\text{for}\quad k=1,\ldots,l,
\end{gather*}
that is, the basis is given by
\begin{gather*}
\bigg(\mathsf{E}(\kappa_2)+\Delta\kappa\frac{\partial}{\partial\kappa_2}\mathsf{E}(\kappa_2) +O\big(\Delta\kappa^2\big),\mathsf{E}(\kappa_2),\mathsf{E}(\kappa_3),\ldots,\mathsf{E}(\kappa_m)\bigg).
\end{gather*}
The $\tau$-function is then given by
\begin{gather}\label{eq:tauD}
\tau(t)=\tau_0(t)+\Delta\kappa\, \tau_1(t)+O\big(\Delta\kappa^2\big),
\end{gather}
where $\tau_0(t)$ is generated by the $(m-1)l$-dimensional basis $\{\mathsf{E}(\kappa_2),\mathsf{E}(\kappa_3),\ldots,\mathsf{E}(\kappa_m)\}$ and the matrix $\tilde A$ is given by the $N\times (m-1)l$ matrix,
\begin{gather}\label{eq:RA}
\tilde A=\begin{pmatrix}
A[n_1]&0& \cdots &0\\
A[n_2]&0&\cdots &0\\
0&A[n_3]&\cdots &0\\
\vdots&\vdots&\ddots&\vdots\\
0&0&\cdots&A[n_m]
\end{pmatrix}
\end{gather}
that is, $\tilde A$ is obtained by adding the first block of $l$ columns to the second block of $l$ columns and removing the first block of $l$ columns. And $\tau_1(t)$ is a $\tau$-function generated by the same matrix~$A$ and the new basis,
\begin{gather}\label{eq:Base1}
\bigg(\frac{\partial }{\partial\kappa_2}\mathsf{E}(\kappa_2), \mathsf{E}(\kappa_2), \mathsf{E}(\kappa_3), \ldots, \mathsf{E}(\kappa_m)\bigg).
\end{gather}

Then we have the following proposition.
\begin{Proposition}\label{prop:Dtau}
If the rank of $\tilde A$ in~\eqref{eq:RA} is less than $N$, then the $\tau$-function generated by the basis given in~\eqref{eq:Base1} and the matrix~$A$ in~\eqref{eq:A1} gives a solution of the $l$-th generalized KdV hierarchy.
\end{Proposition}
\begin{proof}
The rank condition, ${\rm rank}\big(\tilde A\big)<N$, implies $\tau_0(t)=0$. Then from~\eqref{eq:tauD}, we have the limit,
\begin{gather*}
\lim_{\Delta\kappa\to 0}\frac{1}{\Delta\kappa}\tau(t)=\tau_1(t).
\end{gather*}
This means that $\tau_1(t)$ generated by the base~\eqref{eq:Base1} and the matrix~$A$ in~\eqref{eq:A1} is a solution of~the $l$-reduced KP hierarchy.
\end{proof}

 With the rank condition in Proposition~\ref{prop:Dtau}, the following degeneration of the $\kappa$-parameters leads to the change of the basis (see~\eqref{eq:base} below),
\begin{gather*}
 (\mathsf{E}(\kappa_1),\mathsf{E}(\kappa_2),\ldots,\mathsf{E}(\kappa_m))
\xrightarrow[]{\kappa_1\to\kappa_2}\bigg(\frac{\partial}{\partial \kappa_2}\mathsf{E}(\kappa_2),\mathsf{E}(\kappa_2),\ldots,\mathsf{E}(\kappa_m)\bigg) \longrightarrow
\\ \hphantom{ (\mathsf{E}(\kappa_1),\mathsf{E}(\kappa_2),\ldots,\mathsf{E}(\kappa_m))}
\xrightarrow[]{\kappa_2\to\kappa_3}\bigg(\frac{\partial^2}{\partial \kappa_3^2}\mathsf{E}(\kappa_3), \frac{\partial}{\partial\kappa_3}\mathsf{E}(\kappa_3),\mathsf{E}(\kappa_3),\ldots, \mathsf{E}(\kappa_m)\bigg)\longrightarrow \cdots.
\end{gather*}
It is then easy to see that a further confluence of the $\kappa$-parameters leads to the generalized bases~\eqref{eq:gEset}, and appropriate rank conditions on the matrix~$A$ generate a $\tau$-function for the KP hierarchy.

\begin{Remark}
If the matrix~$A$ satisfies the \emph{rank conditions} given in Proposition~\ref{prop:Dtau}, one can use the derivative $\frac{\partial}{\partial \kappa}(\kappa E(t,\kappa))$ instead of $\frac{\partial}{\partial\kappa}E(t,\kappa)$.
Note that the formula~\eqref{eq:gE} is useful for the Schur expansion of the
$\tau$-function as we will see in the later sections. We~denote this generalized base by
\begin{gather}\label{eq:G-base}
\big(\big[\hat{\mathsf{E}}^{(q_1)}\big]_m, \big[\hat{\mathsf{E}}^{(q_2)}\big]_m,\ldots,\big[\hat{\mathsf{E}}^{(q_n)}\big]_m\big),
\end{gather}
where
\begin{gather*}
\big[\hat{\mathsf{E}}^{(q_k)}\big]_m=\big(\hat{\mathsf{E}}^{(q_k)}(\kappa_1),\ldots,
\hat{\mathsf{E}}^{(q_k)}(\kappa_m)\big)\qquad\text{with}\quad
\hat{\mathsf{E}}^{(q_k)}(\kappa_j)=\frac{1}{q_k!}\frac{\partial^{q_k}}{\partial \kappa_j^{q_k}}\big(\kappa_j^{q_k}\mathsf{E}(\kappa_j)\big).
\end{gather*}
Note that $\big[\hat{\mathsf{E}}^{(q_k)}\big]_1=\hat{\mathsf{E}}^{(q_k)}(\kappa_1)$. We~will give several explicit examples in Section~\ref{sec:SGlmk}.
\end{Remark}

\section{The Sato universal Grassmannians}\label{sec:SG}

Here we give a brief description of the Sato universal Grassmannian (see, e.g.,~\cite{Na:18a, SN:84} for further details).
Let~$V=\C((z))$ be the vector space of formal Laurent series,
\begin{gather*}
V:=\C((z))=\bigg\{\sum_{n\in\Z} c_nz^n\colon c_n=0 \text{ for } n\ll 0\bigg\},
\end{gather*}
which has a decomposition,
\begin{gather*}
V=V_{\varnothing}\oplus V_0,
\end{gather*}
where $V_{\varnothing}=\C\big[z^{-1}\big]$ and
$V_0=z\C[[z]]$. Let~$\pi$ be the projection map,
\begin{gather*}
\begin{array}{lccccc}
\pi\colon &V&\longrightarrow &V_{\varnothing},
\\[0.5ex]
&\sum_{n\in\Z}a_nz^n&\longmapsto&\sum_{n\le 0}a_nz^n.
\end{array}
\end{gather*}
Then the Sato universal Grassmannian (UGM) is defined by
\begin{gather*}
\text{UGM}:=\big\{U\subset V\colon \dim\big(\text{Ker}\,\pi\big|_U\big) =\dim \big(\text{Coker}\,\pi\big|_U\big)<\infty\big\}.
\end{gather*}
Each element of UGM is expressed by
\begin{gather*}
U={\rm Span}_{\C}\big\{\phi_{-i}(z)\colon i\in \N_0\big\},
\end{gather*}
where $\N_0=\N\cup\{0\}$. Here the base function $\phi_{-i}(z)$ has the following form: for some
integer $N\gg0$,
\begin{gather}\label{eq:basis}
\phi_{-i}(z)=
\begin{cases}
z^{-i}+\sum\limits_{j> -i}^\infty b_{j,-i}z^j&\text{if}\quad i\ge N,
\\[1.5ex]
\sum\limits_{j> -N}^\infty b_{j,-i}z^j&\text{if}\quad 0\le i\le N-1.
\end{cases}
\end{gather}
The coefficients $\tilde B=(b_{j,-i})$ for $i\in \N_0$ and $j\in\Z$ give the following $\Z\times\N_0$
matrix form, called the \emph{Sato frame} of the element $U\in\text{UGM}$,
\begin{gather*}
\tilde B=\begin{pmatrix}
\ddots & \ddots & & \vdots& \vdots& \vdots & \vdots \\
 & 1 & 0 &0 &\cdots & \cdots &0 \\
\cdots & \vdots &1 & 0 &0 &\cdots &0\\
\cdots & \vdots & \vdots & \young[1][8][$*$] & * &\cdots & *\\
 \cdots & \vdots & \vdots & * & * & \cdots & *\\
 \cdots & \vdots &\vdots &\vdots &\vdots &\cdots &\vdots
\end{pmatrix}\!.
\end{gather*}
Here the columns are labeled by $-\N_0$ from right to left as shown in $(\ldots, -2,-1,0)$, and
the rows are indexed in downward by $\Z$, as in the transpose of $(\ldots, -2,-1,0,1,2,\ldots)$.
The submatrix with $*$'s in the bottom right section corresponds to the coefficients $(b_{j,-i})$ with
$0\le i\le N-1$ and $j\ge -N+1$. Note that the top row with $*$'s is at $j=-N+1$, and
the left most column with $*$'s is at $i=N-1$, that is, the top left corner is at $(-N+1,-N+1)$, marked
by \young[1][6][$*$].

Let~$\hat B$ be a column echelon form of the matrix $\tilde B$. Then write $\hat B$ in the form,
\begin{gather*}
\hat B=[\ldots, B_{-N},B_{-N+1},\dots, B_{-1},B_0].
\end{gather*}
Then we define the set of ordered integers $m=\{m_0>m_1>m_2>\cdots\}$ as
\begin{gather*}
m_{-i}=\begin{cases}
 i&\text{for}\quad i\le -N,
 \\
\min \{j\in\Z\colon b_{j,i}\ne 0\}& \text{for}\quad -N<i\le 0.
\end{cases}
\end{gather*}
Note that $m_{i}$ represents the pivot index of the column $B_{-i}=\{b_{j,-i}\colon j\in\mathbb{Z}\}$ for $i\in\mathbb{N}_0$. And the base functions $\phi_{-i}(z)$ in~\eqref{eq:basis} is expressed by
\begin{gather}\label{eq:phiB}
\phi_{-i}(z)=\big(\ldots,z^{-2},z^{-1},1,z,z^2,\ldots\big) B_{-i}=\sum_{j\in\Z}b_{j,-i}z^{j}.
\end{gather}
The set $m$ is expressed by the so-called \emph{Maya} diagram as follows:
For the case with $m_i=-i$ for all $i\in N_0$ (this is called the \emph{vacuum} state), the
$\hat B$ is a lower triangular matrix whose diagonals are all $1$'s, i.e., $b_{-i,-i}=1$ for all $i\in \mathbb{N}_0$. The corresponding
Maya diagram is then given by the horizontal boxes filled with the black balls as shown below,
\begin{gather*}
\cdots\cdots\young[20][10][\hskip0.3cm\circle*{6},\bb,\bb,\bb,\bb,\bb,\bb,\bb,\bb,\bb]\cdots\cdots,
\end{gather*}
where the index of the box for the right most black ball is ``0'', and the index of each box increases toward right. The vacuum state corresponds to the element with $N=0$. Now taking rightmost $N$ black balls and relocating them so that the boxes filled by the black balls have the index set $m=\{m_0>m_1>\cdots>m_{N-1}>-N>-N-1>\cdots\}$ with
$m_{N-1}>-N+1$, one can represent certain component of UGM. This can be considered as a Schubert decomposition of UGM. For example, the index set $m=(7, 4,2,0,-2,-5,-6,\ldots)$ gives the case $N=5$, which corresponds to
the following Maya diagram,
\begin{gather*}
\cdots\cdots\young[20][10][\hskip0.3cm\circle*{6},\bb,\bb,\bb,\bb,,,\bb, ,\bb,,\bb,,\bb,,,\bb]\cdots\cdots.
\end{gather*}
Here the index of the right most black ball is ``7''.
One should note that the number of filled boxes with the positive indices is the same as that of empty boxes with nonpositive indices. This implies the condition for UGM, i.e., $\dim \big(\text{Ker}|_U\big)=\dim (\text{Coker}|_U)$ for $U\in {\rm UGM}$.

It is also known that the Maya diagram is alternatively expressed by the Young diagram. For~the above example, we have
\begin{center}
\young[7,5,4,3,2][6]
\end{center}
The vertical edges of the southeast boundary of the diagram indicates the index set $m=(m_0,m_1,m_2,\ldots)$: The vertical edge of the top right corner is assigned with
the index $m_0$. Then take the lattice path along the boundary in clockwise direction and
assign each edge with the index starting from $m_0$, so that each vertical edge is labeled by an index from the set $m$. The Young diagram $\lambda=(\lambda_1,\lambda_2,\ldots,\lambda_N)$
is expressed in terms of the index set $m$,
\begin{gather*}
\lambda=(m_0,m_1+1,m_2+2,\ldots,m_{N-1}+N-1).
\end{gather*}
(Note that $m_i-i=0$ for $i\le -N$.)

Using the Young diagrams, we can have the Schubert decomposition of UGM similar to~\eqref{eq:Schubert},
\begin{gather*}
\text{UGM}=\bigsqcup_{\lambda\in\mathcal Y}\tilde X_\lambda,
\end{gather*}
where $\mathcal{Y}$ is the set of all Young diagrams, and each element in the Schubert cell $\tilde X_\lambda$ associated with $\lambda=(m_0,m_1-1,\ldots,m_{k}-k,\ldots)$
with $m_i=-i$ for $i\ge N$ has the form with $b_{j,-i}\in \C$,
\begin{gather*}
{\rm Span}_{\mathbb{C}}\bigg\{\phi_{-i}=z^{m_i}\sum_{j=1}^\infty b_{j,-i}z^j\colon i\in\mathbb{N}_0\bigg\}\in\tilde X_\lambda.
\end{gather*}


\subsection{UGM for the generalized soliton solutions}
Let us identify each soliton solution as a point of UGM. More precisely, we consider an embedding
associated with the matrix $K$ in~\eqref{eq:K} (or $\hat K$ in~\eqref{eq:Kg} for the generalized soliton),
\begin{gather}\label{def:embedding}
\begin{array}{lccccc}
\sigma_K\colon &{\rm Gr}(N,M)&\hookrightarrow &\text{UGM},
\\[0.5ex]
&A &\longmapsto &U={\rm Span}_\C\{\mathcal{B}\},
\end{array}
\end{gather}
where $\mathcal{B}=\{\phi_{-i}(z)\colon i\in\N_0\}$ and each $\phi_{-i}(z)$ is given by
\eqref{eq:phiB}. That is, we have
\begin{gather*}
\phi_{-i}(z)=\big(\ldots,z^{-2},z^{-1}, 1, z, z^2,\ldots\big)\tilde B_{-i}\qquad\text{for}\quad i=0,1,2,\ldots,
\end{gather*}
where $\tilde B_{-i}$ is the $(-i)$-th column of the Sato frame,
\begin{gather*}
\tilde B=\begin{pmatrix}
\ddots & \ddots & & \vdots& \vdots& \vdots & \vdots \\
 & 1 & 0 &0 &0 & \cdots &0 \\
\cdots &0 &1 & 0 &0 &\cdots &0\\
\cdots & 0 & 0 & \young[1][6][$*$] & * &\cdots & *\\
 \cdots & 0 & 0 & * & * & \cdots & *\\
 \cdots & \vdots &\vdots &\vdots &\vdots &\cdots &\vdots
\end{pmatrix}=\big[\ldots, \tilde B_{-2},\tilde B_{-1},\tilde B_0\big].
\end{gather*}
Here the submatrix marked with $*$'s is given by the $\infty\times N$ matrix $KA^{\rm T}$ (or $\hat K A^{\rm T}$), and the top left corner as marked by \young[1][7][$*$] has the coordinates $(-N+1,-N+1)\in \mathbb{Z}\times(-\mathbb{N}_0)$. We~then denote~$\tilde B$ by $\widetilde{KA^{\rm T}}$. Since the basis $\mathcal{B}$ is uniquely determined by the Sato frame $\tilde B$, we sometimes refer to $\mathcal{B}$ as the Sato frame. Recall that our $\tau$-function is given by (2.11) with $\tilde KA^{\rm T}$ embedded into the matrix $\tilde B$, which represents a point of UGM and gives the coefficients in the series solution of the $\tau$-function. We~then have the following proposition.
\begin{Proposition}\label{prop:phi-f}
Each function $\phi_{-i}(z)$ can be expressed by the function $\hat f_{N-i}(t)$ in~\eqref{eq:gfi} in the form,
\begin{gather}\label{eq:f-phi}
\phi_{-i}(z)=c_iz^{-N+1}{\rm e}^{\sum_{n=0}^\infty\frac{1}{n}z^n\partial_n}\hat f_{N-i}(t)\big|_{t=0}\!=c_iz^{-N+1}\hat f_{N-i}([z])\quad\text{for } i\!=\!0,1,\ldots,N\!-\!1,\!\!\!
\end{gather}
where $c_i$ is a normalization constant, and $[z]$ represents
\begin{gather*}
[z]=\bigg(z,\frac{1}{2}z^2,\frac{1}{3}z^3,\ldots\bigg).
\end{gather*}
\end{Proposition}

\begin{proof}
First we have the following formulas (see, e.g.,~\cite{K:17}),
\begin{gather*}
{\rm e}^{\sum_{n=1}^\infty\frac{1}{n}z^n\partial_n}=\sum_{n=0}^\infty z^np_n(\tilde\partial)\qquad\text{and}\qquad
p_n(\tilde\partial)p_m(t)\big|_{t=0}=\delta_{n,m},
\end{gather*}
where $\tilde\partial$ is defined by
\begin{gather*}
\bigg(\partial_1,\frac{1}{2}\partial_2,\frac{1}{3}\partial_3,\ldots\bigg).
\end{gather*}
That is, the operator $\exp\big(\sum \frac{1}{n}z^n\partial_n\big)$ replaces $p_m(t)$ with $z^m$, i.e., \begin{gather}\label{eq:replacement}
{\rm e}^{\sum_{n=1}^\infty\frac{1}{n}z^n\partial_n}p_m(t)\big|_{t=0}=p_m(t+[z])\big|_{t=0}=p_m([z])=z^m.
\end{gather}
 From the formulas of the $\tau$-function~\eqref{eq:tauP} and the function~$\hat f_i(t)$ in~\eqref{eq:gfi},
note that the function~$\phi_{-i}(z)$ is obtained by replacing $p_n(t)$ in~\eqref{eq:gfi} with $z^n$.
Then place $\hat f_1(t)$ at the $(-N+1)$-th column place in the Sato frame, so that
the expansion of $\hat f_1(t)$ starts with the factor $z^{-N+1}$.
This implies the formula in the proposition. The constant $c_i$ normalizes the leading term in
the expansion of $\phi_{-i}(z)$ to be one.
\end{proof}

We note that the exponential functions $\hat E_j^{(q)}([z])$ in~\eqref{eq:gE} is expressed by
\begin{gather}\label{eq:E-Z}
\hat E_j^{(q)}([z])=\frac{1}{q!}\frac{\partial^q}{\partial \kappa_j^q} \bigg(\frac{\kappa_j^q}{1-\kappa_jz}\bigg)=\sum_{n=0}^\infty \binom{q+n}{n}\kappa_j^nz^n,
\end{gather}
where we have used $\sum_{n=1}^\infty\frac{1}{n}\kappa_j^nz^n=-\ln (1-\kappa_jz)$.
 From~\eqref{eq:gfi}, we have the expansion,
\begin{gather*}
\phi_{-i}(z)=c_i z^{-N+1}\sum_{n=0}^\infty\sum_{j=1}^Ma_{i,j}\binom{q_j+n}{n}\kappa_j^nz^n\qquad \text{for}\quad i=0,1,\ldots,N-1.
\end{gather*}


In conclusion, the embedding $\sigma_{\hat K}$ for the generalized soliton solution gives
$\sigma_{\hat K}(A)={\rm Span}_\C\{\mathcal{B}\}$ $\in \text{UGM}$ with a basis $\mathcal{B}\subset \C((z))$,
\begin{gather*}
\mathcal{B}=\big\{\phi_{-i}(z)=c_iz^{-N+1}\hat f_{N-i}([z]) \ (0\le i\le N-1),\ \phi_{-i}(z)=z^{-i}\ (i\ge N)\big\},
\end{gather*}
where $c_i$'s are chosen so that each $\phi_{-i}(z)$ is a monic function in $\C((z))$. We~also remark that $\hat f_{i}([z])$ is a rational function with some power of $(1-(\kappa_jz)^l)^{-1}$, see~\eqref{eq:gfi} and~\eqref{eq:E-Z}. Then multiplying $\mathcal{B}$ by $\prod_{j=1}^m\big(1-(\kappa_jz)^l\big)^{q_j}$, we have
\begin{gather*}
\hat{\mathcal{B}}:=\bigg(\prod_{j=1}^m\big(1-(\kappa_jz)^l\big)^{q_j}\bigg) \mathcal{B}\subset\C\big[z,z^{-1}\big],
\end{gather*}
that is, each element of $\hat{\mathcal{B}}$ is a Laurent polynomial. We~further multiply $\hat{\mathcal{B}}$ by $z^{-g}$ for some positive integer $g$, so that we have a subset of polynomials in $\C\big[z^{-1}\big]$ including 1, i.e.,
\begin{gather*}
{\mathcal{A}}=z^{-g}\hat{\mathcal{B}}\subset \C\big[z^{-1}\big]\qquad\text{and}\qquad 1\in\mathcal{A},
\end{gather*}
which is a central object in this paper. Then the main theorem in this paper (Theorem~\ref{thm:Main}) shows that the subset $\mathcal{A}$ for a certain class of soliton solutions gives a subalgebra in $\C[z^{-1}]$.

\begin{Remark}
The subset $\hat{\mathcal{B}}$ also gives a point of UGM, which is different from the point given by $\mathcal{B}$. However, the solution $u$ of the KP hierarchy remains the same for both $\mathcal{B}$ and $\hat{\mathcal{B}}$. This can be seen as follows. Writing $\prod_{j=1}^m\big(1-(\kappa_jz)^l\big)^{q_j}$ in the exponential form, i.e.,
\begin{gather*}
\prod_{j=1}^m(1-(\kappa_jz)^l)^{q_j}
=\exp\bigg(\sum_{j=1}^mq_j\ln\big(1-(\kappa_jz)^l\big)\bigg): =\exp\bigg(\sum_{n=1}^\infty\frac{1}{n}\alpha_nz^{nl}\bigg),
\end{gather*}
we have the following formula for the function $\hat g_i(t):=\hat f_i(t)\exp\big(\sum_{n=1}^\infty\alpha_nt_{nl}\big)$,
\begin{gather*}
{\rm e}^{\sum_{n=1}^\infty\frac{1}{n}z^n\partial_n}\,\hat g_i(t)\Big|_{t=0}=\hat f_i([z])\exp\bigg(\sum_{n=1}^\infty\frac{1}{n}\alpha_nz^{nl}\bigg).
\end{gather*}
That is, we use $\hat g_i(t)$ for the $\tau$-function instead of $\hat f_i(t)$. Then the $\tau$-function from the functions $\{\hat g_i(t)\colon i=1,\ldots,N\}$ becomes
\begin{gather*}
\tau=\text{Wr}(\hat g_1,\ldots, \hat g_N)={\rm e}^{N\sum_{n=1}^\infty\alpha_nt_{nl}}\text{Wr}(\hat f_1,\ldots,\hat f_N),
\end{gather*}
which implies that the solution $u=2\partial_1^2\ln \tau$ is invariant under this multiplication (also see~\cite{Ta:89}).
\end{Remark}

\subsection[The soliton solutions of the l-th generalized KdV hierarchy]{The soliton solutions of the $\boldsymbol l$-th generalized KdV hierarchy}

We here consider the following simple but useful example of the soliton solution associated with~${\rm Gr}(1,l)$. From~\eqref{eq:reduction}
and~\eqref{eq:f}, we have
$r=\kappa^l$, and the $\kappa$-parameters are given by
\begin{gather*}
(\kappa_1,\kappa_2,\ldots,\kappa_l)=\big(\kappa,\omega_l\kappa,\omega_l^2\kappa,\ldots, \omega_l^{l-1}\kappa\big)\qquad\text{with}\quad \omega_l=\exp\bigg(\frac{2\pi {\rm i}}{l}\bigg).
\end{gather*}
We take the matrix $A=\big(1,\omega_l,\omega_l^2,\ldots,\omega_l^{l-1}\big)\in {\rm Gr}(1,l)$ and the $\infty\times l$ Vandermonde matrix $K$ in~\eqref{eq:K},
\begin{gather*}
K=\begin{pmatrix}
1&1&\cdots &1&\\
\kappa&\omega_l\kappa&\cdots &\omega_l^{l-1}\kappa\\
\kappa^2&\omega_l^2\kappa^2&\cdots &\omega_l^{l-2}\kappa^2\\
\kappa^3&\omega_l^3\kappa^3&\cdots &\omega_l^{l-3}\kappa^3\\
\vdots &\vdots &\vdots &\vdots \\
\kappa^{l-1}&\omega_l^{l-1}\kappa^{l-1}&\cdots&\omega_l\kappa^{l-1}\\
\kappa^l &\kappa^l &\cdots &\kappa^l\\
\vdots &\vdots&\vdots&\vdots
\end{pmatrix}\!.
\end{gather*}
The following notations are useful: Let~$\Omega_l^k$ be a vector given by
\begin{gather*}
\Omega_l^k:=\big(1,\omega_l^k,\omega_l^{2k},\ldots,\omega_l^{k(l-1)}\big)\qquad
\text{for}\quad 0\le k\le l-1,
\end{gather*}
which satisfy the orthogonality relations,
\begin{gather*}
\Omega_l^i\cdot\Omega_l^j=l\delta_{i+j,l}\qquad\text{for}\quad 0\le i,j\le l-1.
\end{gather*}
Let us take the matrix~$A$ as $A=\Omega_l^1$, i.e.,
\begin{gather*}
A=\big(1,\omega_l,\omega_l^2,\ldots,\omega_l^{l-1}\big).
\end{gather*}
Then we have
\begin{gather*}
KA^{\rm T}=\big(\underbrace{0,\ldots,0,\,l\kappa^{l-1}}_{l}\,,\underbrace{0,\ldots,0,\, l\kappa^{2l-1}}_{l},\,0,\ldots\big)^{\rm T},
\end{gather*}
which leads to the expansion form of the $\tau$-function,
\begin{gather*}
\tau(t)=f_1(t)=\sum_{i=1}^l \omega_l^{i-1}\exp\bigg(\sum_{n=1}^\infty (\kappa\omega_l^{i-1})^nt_n\bigg)=l\sum_{n=1}^\infty \kappa^{nl-1}p_{nl-1}(t).
\end{gather*}
Note that the corresponding soliton solution is singular, and the leading Schur function, $p_{l-1}(t)$,
has the Young diagram with $l-1$ boxes in horizontal.

Under the embedding $\sigma_K$ in~\eqref{def:embedding}, the image $\sigma_K(A)$ gives
the basis $\{\phi_i(z)\colon -i\in\mathbb{N}_0\}$ in~\eqref{eq:basis},
\begin{gather*}
\mathcal{B}=\bigg\{\phi_0(z)=\frac{z^{l-1}}{1-(\kappa z)^l},\text{ and } \phi_{-i}(z)=z^{-i}\ (i\ge 1)\bigg\}.
\end{gather*}
Here we have used $\phi_0(z)$ in~\eqref{eq:f-phi} with the formula~\eqref{eq:replacement} and the normalization $c_0=\big(l\kappa^{l-1}\big)^{-1}$, i.e.,
\begin{gather*}
\phi_0(z)=c_0f_1([z])=c_0l\sum_{n=1}^\infty (\kappa z)^{nl-1}=c_0\frac{l(\kappa z)^{l-1}}{1-(\kappa z)^l}.
\end{gather*}
Multiplying $\mathcal{B}$ by $z^{-l+1}\big(1-(\kappa z)^l\big)$, we have a basis of a subspace of the set of functions having pole only at $z=0$,
\begin{gather*}
{\mathcal{A}}=\big\{1,z^{-l}, z^{-i-l}\big(1-(\kappa z)^l\big) (i\ge 1)\big\}\subset \C\big[z^{-1}\big].
\end{gather*}

We then note that the set $\mathcal{A}$ represents a commutative ring of polynomials defined on
a~certain (singular) algebraic curve. In order to show this, we introduce the following variables,
\begin{gather}\label{eq:coordinates}
x=z^{-l},\qquad
\text{and}\qquad
y_i=z^{-l-i}\big(1-(\kappa z)^l\big)=z^{-i}\big(x-\kappa^l\big)\qquad
\text{for}\quad i=1,\ldots,l-1,
\end{gather}
Then $\mathcal{A}$ gives a set of meromorphic functions having pole only at $\infty$
in these variables,
\begin{gather}\label{eq:Alg}
\tilde{\mathcal{A}}=\big\{1,x,y_1,\ldots,y_{l-1},x^2,xy_1,\ldots,xy_{l-1},x^3,x^2y_1,\ldots\big\}.
\end{gather}
Note here that we have the following relations among the variables $(x,y_1,\ldots,y_{l-1})$,
\begin{gather*}
p_{i,j}:=y_iy_{j-1}-y_jy_{i-1}=0\qquad\text{for}\quad 1\le i<j\le l,
\end{gather*}
where we define $y_0=F(x)=x-\kappa^l$ and $y_l=G(x)=xF(x)$. These $p_{i,j}$ can be expressed as the $2\times 2$ minors of the $2\times l$ matrix,
\begin{gather}\label{eq:2xl}
\begin{pmatrix}
y_1 &y_2 &\cdots &y_{l-1}& G(x)
\\
F(x)&y_1&\cdots &y_{l-2}&y_{l-1}
\end{pmatrix}\!.
\end{gather}
This expression will be very useful when we consider the deformation of the singular curves
associated with these relations in Section~\ref{sec:deformation}.

It is then important to notice that the vector space generated by the set $\tilde{\mathcal{A}}$ in~\eqref{eq:Alg} can be expressed by a commutative ring $\mathcal{R}$ of the variables $(x,y_1,\ldots,y_{l-1})$, i.e.,
\begin{gather*}
\RR=\C[x,y_1,y_2,\ldots,y_{l-1}]/\mathcal{P}\cong{\rm Span}_\C\{\tilde{\mathcal{A}}\},
\end{gather*}
where the prime ideal is given by the minors,
\begin{gather*}
\mathcal{P}:=\big\{p_{i,j}=0\colon 1\le i<j\le l\big\}.
\end{gather*}
That is, the variables $(x,y_1,\ldots,y_{l-1})$ are the coordinates for a singular curve $\mathcal{C}$, whose affine part is given by $\text{Spec}(\RR)$.

We also note that the relations $\{p_{i,j}(x,y_1,\ldots,y_{l-1})=0\}$ define an irreducible component of the intersections of the (singular) hypersurfaces,
\begin{gather*}
y_i^l=G(x)^iF(x)^{l-i}=x^i\big(x-\kappa^l\big)^l\qquad\text{for}\quad i=1,\ldots,l-1,
\end{gather*}
which will be discussed further in Section~\ref{sec:deformation}.

We note that the ring $\RR$ is graded with the degree of the polynomials and has a \emph{numerical semigroup} structure of type,
\begin{gather*}
\langle l,l+1,\ldots, 2l-1\rangle,
\end{gather*}
which gives the following table for the monomials in~\eqref{eq:Alg},
\begin{center}\renewcommand{\arraystretch}{1.2}
\begin{tabular}{c| c| c| c| c| c| c| c| c| c| c| c| c c c}
\hline
0&1&$2$ &$\cdots$&$l-1$ &$l$&$l+1$ &$\cdots$ &$2l-1$& $2l$ &$2l+1$&$2l+2$&$\cdots$ \\ \hline
1&$*$&$*$&$\cdots$&$*$ &$x$&$y_1$&$\cdots$&$y_{l-1}$&$x^2$ & $xy_1$ &$xy_2$&$\cdots$\\
\hline
\end{tabular}
\end{center}
where $*$'s are the gaps of the sequence $(0,1,2,\ldots)$. The number of gaps is called (arithmetic) genus of the numerical semigroup. In Section~\ref{sec:NS}, we describe the numerical semigroup of type $\langle l,lm+1,lm+2,\ldots,lm+k\rangle$ for any $l\ge 2$, $m\ge 1$ and $1\le k\le l-1$. One should note in general that the geometric genus is not the same as the arithmetic one for the singular curve.

\section[The numerical semigroups for the generalized soliton solutions]{The numerical semigroups\\ for the generalized soliton solutions}\label{sec:NS}

Let us start with a brief introduction to the numerical semigroups (see, e.g.,~\cite{RG:09} for the details).

\subsection{The numerical semigroups}
A numerical semigroup $S$ is a subset of $\mathbb{N}_0$, such that it
\begin{itemize}\itemsep=0pt
\item[(1)] contains $0$, i.e., $0\in S$,
\item[(2)] is closed under addition, i.e., for any $a,b\in S$, $a+b\in S$, and
\item[(3)] has a finite complement in $\mathbb{N}_0$, i.e., $|\mathbb{N}_0\setminus S|<\infty$.
\end{itemize}
The finite set $\mathbb{N}_0\setminus S$ is called \emph{gap sequence} or \emph{gaps} of $S$.

It is known that every numerical semigroup is generated by a finite set $\{x_1,x_2,\ldots, x_n\}\subset \mathbb{N}$, i.e.,
\begin{gather*}
S=\langle x_1, x_2,\ldots, x_n\rangle:=\bigg\{\!\sum_{i=1}^n a_ix_i\colon a_i\in\mathbb{N}_0\bigg\}.
\end{gather*}
It is also well known that $S$ is a numerical semigroup if and only if $\gcd\{x_1,\ldots,x_n\}=1$.
Every numerical semigroup has a unique finite set which generates $S$. We~also have the following notions for the numerical semigroup.
\begin{itemize}\itemsep=0pt
\item[$(a)$] The (arithmetic) genus of $S$ is $g(S)=|\mathbb{N}_0\setminus S|$.
\item[$(b)$] The Frobenius number $F(S)$ is the finite greatest positive integer in $\mathbb{N}_0\setminus S$.
\end{itemize}

A numerical semigroup can be expressed by a Young diagram, denoted by $\lambda(S)$ as follows. The southeast boundary
of the Young diagram gives a path labeled by the numbers from $0\in S$ starting at the northeast corner, so that the elements in $S$ appear on the vertical edges, hence the gaps on the horizontal edges, in the boundary path as shown below.
\setlength{\unitlength}{0.5mm}
\begin{center}
 \begin{picture}(160,60)
\put(5,55){\line(1,0){110}}
\put(5,45){\line(1,0){110}}
 \put(5,35){\line(1,0){90}}
 \put(5,25){\line(1,0){90}}
 \put(5,15){\line(1,0){70}}
 \put(5,5){\line(1,0){70}}
 \put(5,-20){\line(0,1){75}}
 \put(31,5){\line(0,1){50}}
 \put(18,5){\line(0,1){50}}
 \put(75,5){\line(0,1){10}}
 \put(62,5){\line(0,1){10}}
 \put(95,25){\line(0,1){10}}
 \put(82,25){\line(0,1){10}}
\put(115,45){\line(0,1){10}}
\put(102,45){\line(0,1){10}}
 \put(98,29){${s_i}$}
 \put(65,37){$\vdots$}
 \put(10,37){$\vdots$}
 \put(24,37){$\vdots$}
 \put(60,48){$\cdots$}
 \put(55,28){$\cdots$}
 \put(47,17){$\vdots$}
 \put(10,17){$\vdots$}
 \put(24,17){$\vdots$}
 \put(45,8){$\cdots$}
 \put(78,8){$s_j$}
 \put(118,48){$0$}
 \put(10,-18){$\vdots$}
 \put(5,-3){$F(S)\quad\cdots\cdots\cdots\quad$}
 \put(60,-3){$ \hskip0.2cm g_m$}
 \end{picture}
\end{center}
\vskip1cm

\noindent
Here $0, s_i, s_j \in S$ and $ g_m, F(S)\in \mathbb{N}_0\setminus S$. Note that the Young diagram
$\lambda=(\lambda_1,\ldots,\lambda_N)$ is expressed by $\lambda_k=F(S)-N+k-s_{k-1}$ with $s_0=0$.

\subsection[The numerical semigroups of type <l,lm+1,...,lm+k>]{The numerical semigroups of type $\boldsymbol{\langle l,lm+1,\ldots,lm+k\rangle}$}

In this paper, we consider the numerical semigroups of type $\langle l,lm+1,lm+2,\ldots,lm+k\rangle$,
where $l\ge 2, m\ge 1$ and $1\le k\le l-1$.

\begin{Proposition}\label{prop:FG}
The Frobenius number $F(S)$ and the genus $g(S)$ of the numerical semigroup $S=\langle l,lm+1,lm+2,\ldots,lm+k\rangle$ are given by
\begin{gather*}
F(S)=mln_{l,k}-1,\qquad g(S)=mn_{l,k}\bigg(l-1-\frac{k}{2}(n_{l,k}-1)\bigg),
\end{gather*}
where ${n_{l,k}=\big\lceil\frac{l-1}{k}\big\rceil}$.
\end{Proposition}

\begin{proof}
Let us first consider the case of $m=1$.
Since $l$ is the smallest generator of $S$ and $l+k$ is the largest one, there exists
the largest number denoted by $n_{l,k}$ such that $l(n+1)-(l+k)n>1$ for $n\le n_{l,k}$.
This largest number $n_{l,k}$ is given by
$n_{l,k}=\big\lceil \frac{l-1}{k}\big\rceil$.
This means that any number $a\ge ln_{n,k}$ is in $S$, and
we have $F(S)=ln_{l,k}-1$.

Let~$[a,b]$ denote the set of numbers $\{a,a+1,\ldots,b\}$ for $a<b\in\N_0$.
Since the gaps are the integers in the set,
\begin{gather*}
[1,l-1]\cup[l+k+1,2l-1]\cup\cdots\cup[(l+k)(n_{l,k}-1)+1,ln_{l,k}-1],
\end{gather*}
the number of gaps are counted by
\begin{gather*}
g(S)=\sum_{n=0}^{n_{l,k}-1}(l-1-kn)=n_{l,k}\bigg(l-1-\frac{k}{2}(n_{l,k}-1)\bigg).
\end{gather*}

For the case with $m>1$, we have $m$ copies of the above situation and obtain the formulas asserted in the proposition.
\end{proof}

\subsubsection{Young diagrams}

The generalized soliton solution associated with the semigroup $S=\langle l,lm+1,lm+2,\ldots,lm+k\rangle$ is generated from a point of Grassmannian ${\rm Gr}(N,M)$ with
\begin{gather*}
M=F(S)+1,\qquad N=M-g(S).
\end{gather*}
This is easy to see from the Young diagram given in the previous section.
Also from the proof of Proposition~\ref{prop:FG}, we have that
for $m=1$, the genus $g(S)$ has a partition,
\begin{gather*}
g(S)=\sum_{i=1}^{n_{l,k}}g_i(S)\qquad\text{with}\quad g_i(S)=l-1-k(i-1).
\end{gather*}

The following lemma is easy to prove.
\begin{Lemma}
For $m=1$, the length $N=N(S)$ of the Young diagram $\lambda(S)$ has a partition,
\begin{gather*}
N(S)=\sum_{i=1}^{n_{l,k}}n_i(S)\qquad\text{with}\quad n_i(S)=k(i-1)+1.
\end{gather*}
Note that $n_i(S)+g_i(S)=l$.
\end{Lemma}
\begin{proof}
Let~$I_i$ be the set $[il,(i+1)l-1]$. Then we have $(I_i\cap S)\cup(I_i\cap G(S))=I_i$,
where $G(S)$ is the set of gaps in $S$. This implies the lemma.
\end{proof}

Then the Young diagram $\lambda(S)$ for $S=\langle l,l+1,l+2,\ldots,l+k\rangle$ ($m=1$) has the
following shape: each rectangular box (shaded region in the figure below) at the southeast boundary has the size $g_i(S)\times n_i(S)$,
i.e., $g_i(S)$ boxes in horizontal, and $n_i(S)$ in vertical, as shown below.

\begin{figure}[h]
\centering
\includegraphics[height=5.8cm]{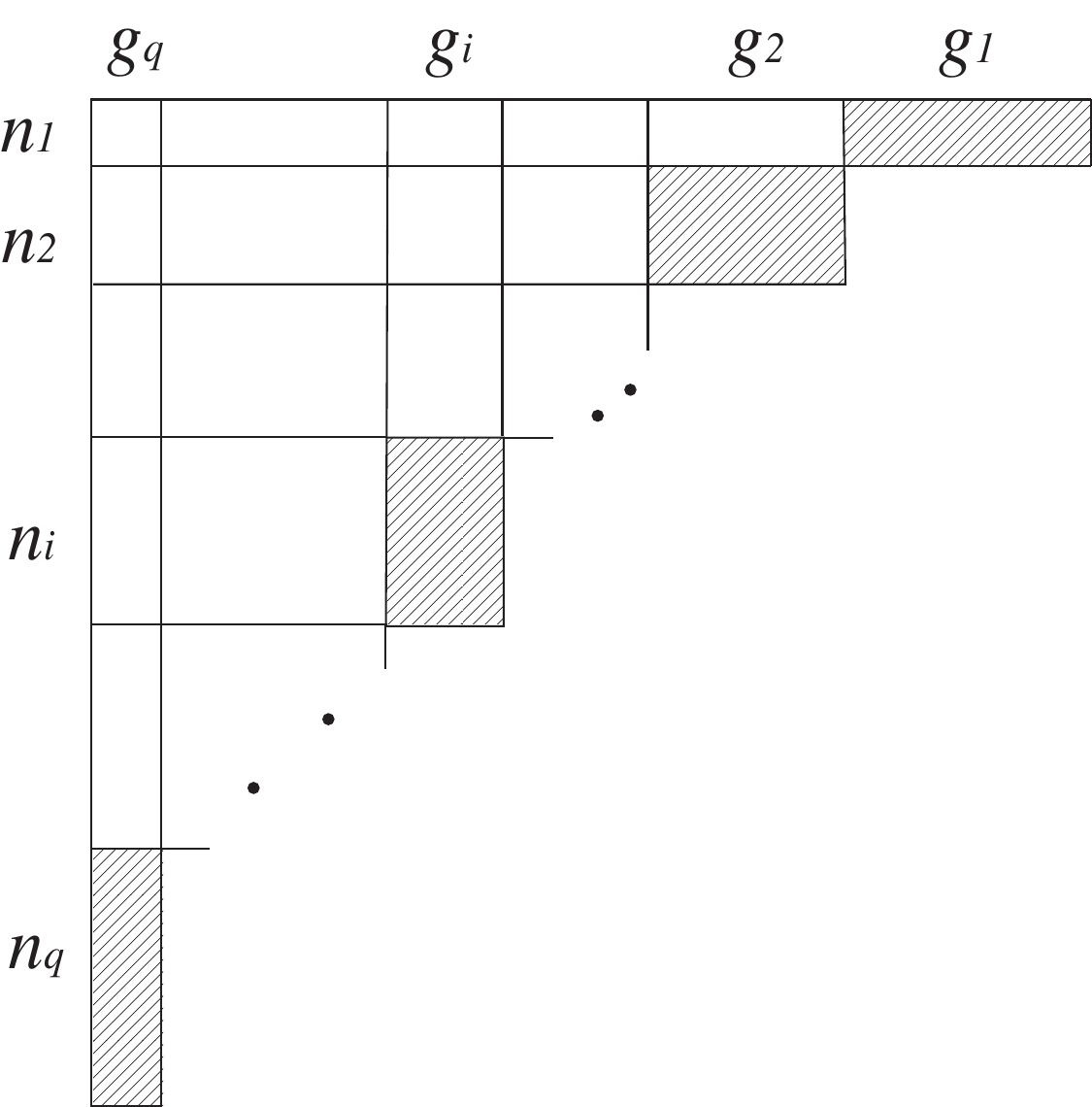}
\end{figure}

In the figure, $q$ in the suffix represents $q=n_{l,k}=\big\lceil\frac{l-1}{k}\big\rceil$.

For general case $m\ge1$, the Young diagram consists of the partitions of the horizontal direction,
\begin{gather*}
(\underbrace{g_1,\ldots,g_1}_{m},\underbrace{g_2,\ldots,g_2}_m,\ldots,\underbrace{ g_{n_{l,k}},\ldots,g_{n_{l,k}}}_m).
\end{gather*}
That is, the Young diagram $\lambda(S)=(\lambda_1,\lambda_2, \ldots,\lambda_N)$ is given by
\begin{gather*}
\lambda_{i_k}=m\sum_{j=k}^{n_{l,k}}g_j\qquad\text{for}\quad N_{k-1}\le i_k\le N_k,
\end{gather*}
where $N_k=\sum_{i=1}^kn_i$ for $k=1,\ldots,n_{l,k}$ and $N=mN_{n_{l,k}}$.
Likewise, in the vertical direction, we have
\begin{gather*}
(\underbrace{n_1,\ldots,n_1}_{m},\underbrace{n_2,\ldots,n_2}_m,\ldots,\underbrace{ n_{n_{l,k}},\ldots,n_{n_{l,k}}}_m),
\end{gather*}
where $g_i+n_i=l$ for all $i=1,\ldots,n_{l,k}$.

\section[The generalized solitons associated with the semigroup [S=<l,lm+1,\ldots,lm+k>]{The generalized solitons associated\\
 with the semigroup $\boldsymbol{S=\boldsymbol \langle l,lm+1,\ldots,lm+k\rangle}$}\label{sec:SGlmk}

We take the base functions~\eqref{eq:G-base} with the following set of exponentials,
\begin{gather}\label{eq:base}
\big( \big[\hat{\mathsf{E}}^{(q-1)}\big]_m,\big[\hat{\mathsf{E}}^{(q-2)}\big]_m,\ldots,\big[\hat{\mathsf{E}}^{(0)}\big]_m\big)\qquad
\text{with}\quad q=n_{l,k}=\bigg\lceil\frac{l-1}{k}\bigg\rceil,
\end{gather}
where
\begin{gather*}
[\hat{\mathsf{E}}^{(j)}]_m=\big(\hat{\mathsf{E}}^{(j)}(\kappa_1),\ldots, \hat{\mathsf{E}}^{(j)}(\kappa_m)\big)\qquad\text{with}\quad \hat{\mathsf{E}}^{(j)}(\kappa_i)=\frac{1}{j!}\frac{\partial^{j}}{\partial \kappa_i^{j}}\big(\kappa_i^{j}\mathsf{E}(\kappa_i)\big),
\end{gather*}
and $\mathsf{E}(\kappa_i)=(E_1(t,\kappa_i),\ldots, E_l(t,\kappa_i))$ with $E_p(t,\kappa_i)=\exp\big(\sum_{n=1}^\infty \big(\kappa_i\omega_l^{p-1}\big)^nt_n\big)$.
Thus, the base functions~\eqref{eq:base} span an $lmq$-dimensional space.

{\samepage With the expansion $\kappa^jE(t, \kappa)=\sum_{n=0}^\infty\kappa^{j+n}p_n(t)$, we have
\begin{gather*}
\hat{\mathsf{E}}^{(j)}(\kappa_i)=(1,p_1(t), p_2(t),\ldots)\hat K_i^{(j)},
\end{gather*}
where $\hat K_i^{(j)}$ is an $\infty\times l$ matrix whose $n$-th row (counting from $n=0$) is the $l$-dimensional vector given by
\begin{gather}\label{eq:Kj}
\text{Row}_n(\hat K_i^{(j)})=\binom{n+j}{j}\kappa_i^{n}\Omega_l^n,
\end{gather}
with $\Omega_l^n=\big(1,\omega_l^n,\omega_l^{2n},\ldots,\omega_l^{n(l-1)}\big)$.

}

Then the $\tau$-function~\eqref{eq:tauP} of the generalized soliton solution
of the $l$-reduction is given by
\begin{gather}\label{eq:tauMain}
\tau(t)=\left|
\begin{pmatrix}
1& p_1&p_2&\cdots&\cdots &\cdots
\\
0& 1 & p_1 &p_2&\cdots&\cdots
\\
\vdots&\ddots&\ddots &\ddots &\ddots&\vdots
\\
0&\cdots& 0&1&p_1&\cdots
\end{pmatrix}\hat K A^{\rm T}\right|,
\end{gather}
where $\hat K$ is given by an $\infty\times lmq$ matrix,
\begin{gather*}
\hat K=\big(\big[\hat K^{(q-1)}\big]_m,\big[\hat K^{(q-2)}\big]_m,\ldots,\big[\hat K^{(0)}\big]_m\big),
\end{gather*}
with the $\infty\times lm$ matrix,
\begin{gather*}
[\hat K^{(j)}]_m:=\big(\hat K_1^{(j)},\hat K^{(j)}_2,\ldots,\hat K_m^{(j)}\big).
\end{gather*}

Then we have the following main theorem in this paper.
\begin{Theorem}\label{thm:Main}
Let~$(l,m,k)$ be a triplet of numbers satisfying $l\ge 2, m\ge 1$ and $1\le k\le l-1$.
Consider a generalized soliton solution based on the exponential functions~\eqref{eq:base} with $q=n_{l,k}=\big\lceil\frac{l-1}{k}\big\rceil$ and the $A$-matrix given by
\begin{gather}\label{eq:Amatrix}
A=\begin{pmatrix}
I_m\otimes [\Omega_l]^{n_1} & 0&\cdots & 0\\
0 & I_m\otimes [\Omega_l]^{n_2}&\cdots &0\\
\vdots &\ddots &\ddots &\vdots\\
0 &\cdots &0&I_m\otimes [\Omega_l]^{n_q}
\end{pmatrix}\!,
\end{gather}
where $I_m$ is the $m\times m$ identity matrix, $n_i=k(i-1)+1$, and $[\Omega_l]^n$ is an $n\times l$ matrix defined by
\begin{gather*}
[\Omega_l]^n=\begin{pmatrix}
\Omega_l^1 \\ \Omega_l^2 \\ \vdots \\ \Omega_l^n
\end{pmatrix}\qquad\text{with}\quad \Omega_l^k=\big(1,\omega_l^k,\omega_l^{2k},\ldots,\omega_l^{(l-1)k}\big).
\end{gather*}
Then we have the followings:
\begin{itemize}\itemsep=0pt
\item[$(a)$] The matrix~$A$ belongs to a Schubert cell $X_\lambda\subset {\rm Gr}(N,M)$, where
\begin{gather*}
N=mq\bigg(1+\frac{k}{2}(q-1)\bigg)\qquad\text{and}\qquad M=mlq,
\end{gather*}
and the Young diagram $\lambda$ is given by
\begin{gather*}
\lambda=([\lambda_{1,1}]_{n_1},\ldots,[\lambda_{1,m}]_{n_1},[\lambda_{2,1}]_{n_2},\ldots,
[\lambda_{2,m}]_{n_2},\ldots,[\lambda_{q,1}]_{n_q},\ldots,[\lambda_{q,m}]_{n_q}),
\end{gather*}
with
\begin{gather*}
[\lambda_{a,b}]_{n_a}=(\underbrace{\lambda_{a,b},\ldots,\lambda_{a,b}}_{n_a})\qquad\text{for}\quad \begin{cases}
1\le a\le q,\\
1\le b\le m.
\end{cases}
\end{gather*}
Here $\lambda_{a,b}$ is
\begin{gather*}
\lambda_{a,b}=(m-b+1)g_a+mg_{a+1}+\cdots +mg_q,
\end{gather*}
with $g_a=l-n_a=l-1-k(a-1)$.
\item[$(b)$] The image of the embedding $\sigma_{\hat K}(A)\in \text{UGM}$ with $\hat K$ in~\eqref{eq:Kj} can be parametrized by
 a~numerical semigroup of type $S=\langle l,lm+1,lm+2,\ldots,lm+k\rangle$ with the Young diagram
 $\lambda(S)=\lambda$.
\item[$(c)$] The generalized soliton solution generated by the $\tau$-function~\eqref{eq:tauMain} is then associated with an irreducible component of the following singular space curve,
\begin{gather*}
\mathcal{C}=\bigg\{(x,y_1,\ldots,y_{k})\in\C^{k+1}\colon y_i^l=x^i
\prod_{j=1}^m(x-\kappa_j^l)^l\text{ for } 1\le i\le k\bigg\}.
\end{gather*}
\item[(d)]
The $\tau$-function of~\eqref{eq:tauMain} has the following Schur expansion,
\begin{gather*}
\tau(\hat t)=S_\lambda(\hat t)+\sum_{\mu\supset\lambda}c_{\hat\mu}(\kappa)S_\mu(\hat t),
\end{gather*}
where $\hat t=\{t_n\colon n\not\equiv 0\,({\rm mod}\,l)\}$ and $c_{\hat\mu}(\kappa)$ is a homogeneous symmetric polynomial of $\{\kappa_i^l\colon i=1,\ldots,m\}$. Here the Young diagram $\lambda$ is given by $\lambda(S)$
of the numerical semigroup~$S$, and each coefficient $c_{\hat\mu}$ is uniquely determined by $\mu$.
\end{itemize}
\end{Theorem}
We give the proof of Theorem~\ref{thm:Main} in Section~\ref{sec:Proof}.
In the next subsections, we give some examples to illustrate the theorem.

\subsection{2-reductions: KdV reduction}
In this case, we have $S=\langle 2,2m+1\rangle$, i.e., $l=2$ and $k=1$, which implies $n_{2,1}=1$.
The~Fro\-benius number and genus are
\begin{gather*}
F(S)=2m-1,\qquad\text{and}\qquad g(S)=m.
\end{gather*}
The Young diagrams for $m=1,2,3$ are given by
\begin{center}
\young[1][6]\!\!,\qquad \young[2,1][6]\!\!,\qquad \young[3,2,1][6]\!\!.
\end{center}
The soliton solution we consider here has the following $\kappa$-parameters and the
$m\times 2m$ matrix $A\in {\rm Gr}(m,2m)$,
\begin{gather*}
\big(\kappa_1\Omega_2^1,\kappa_2\Omega_2^1,\ldots,\kappa_m\Omega_2^1\big)\qquad
\text{and}\qquad
A=I_m\otimes \Omega_2^1=
\begin{pmatrix}
\Omega_2^1 &0_2 &\cdots &0_2\\
0_2 &\Omega_2^1 &\cdots &0_2\\
\vdots &\vdots&\ddots&\vdots\\
0_2&\cdots &\cdots&\Omega_2^1
\end{pmatrix}\!,
\end{gather*}
where $\Omega_2^1=(1,-1)$ and $0_2=(0,0)$.
The matrix $K$ is an $\infty\times 2m$ matrix,
\begin{gather*}
K=\begin{pmatrix}
1_2&1_2&\cdots &1_2\\
\kappa_1\Omega_2^1&\kappa_2\Omega_2^1&\cdots&\kappa_m\Omega_2^1\\[.3ex]
\kappa_1^2 1_2 &\kappa_2^2 1_2 &\cdots &\kappa_m^2 1_2\\[.3ex]
\kappa_1^3\Omega_2^1&\kappa_2^3\Omega_2^1&\cdots&\kappa_m^3\Omega_2^1\\
\vdots &\vdots&\cdots &\vdots
\end{pmatrix}\!,
\end{gather*}
where $1_2=\Omega_2^0=(1,1)$.
Then the $\infty\times m$ matrix $KA^{\rm T}$ becomes
\begin{gather*}
KA^{\rm T}=\begin{pmatrix}
0 & 0& \cdots & 0\\
2\kappa_1&2\kappa_2&\cdots &2\kappa_m\\
0& 0 & \cdots & 0\\
2\kappa_1^3 &2\kappa_2^3&\cdots &2\kappa_m^3\\
\vdots &\vdots &\vdots &\vdots
\end{pmatrix} \equiv
\begin{pmatrix}
0 & 0& \cdots & 0\\
1&1&\cdots &1\\
0& 0 & \cdots & 0\\
\kappa_1^2 &\kappa_2^2&\cdots &\kappa_m^2\\
\vdots &\vdots &\vdots &\vdots
\end{pmatrix}\!.
\end{gather*}
Through the embedding $\sigma_K$ in~\eqref{def:embedding}, the image $\sigma_K(A)$ gives
the basis $\{\phi_{-i}\colon i\in\N_0\}$ in~\eqref{eq:basis},
\begin{gather*}
\mathcal{B}=\bigg\{\phi_{-i+1}(z)=\frac{z^{-m+2}}{1-(\kappa_iz)^2}\ (\text{for }i=1,\ldots,m),\ \phi_{-j}(z)=z^{-j}\ (\text{for }j\ge m)\bigg\}.
\end{gather*}
We then normalize the set by multiplying
$z^{-g}\prod_{i=1}^m\big(1-(\kappa_iz)^2\big)$ with $g=m$. Then we have a~basis of a subspace in $\C\big[z^{-1}\big]$,
\begin{gather*}
\bigg\{\!
z^{-2m+2}\prod_{p\ne i}^m\!\big(1-(\kappa_pz)^2\big)\, (\text{for } i=1,\ldots,m),\,
z^{-j-m}\prod_{i=1}^m\!\big(1-(\kappa_iz)^2\big)\, (\text{for}\, j\ge m)\!\bigg\}\subset \C\big[z^{-1}\big].
\end{gather*}
Note that the first $m$ terms can be expressed simply by the set
$\big\{1, z^{-2},\ldots,z^{-2m+2}\big\}$. Then this basis can be expressed in a simple form,
\begin{gather*}
\mathcal{A}=\bigg\{1, z^{-2},\ldots,z^{-2m+2},z^{-j-m}\prod_{i=1}^m\big(1-(\kappa_iz)^2\big)\
(\text{for }j\ge m)\bigg\}.
\end{gather*}
We then introduce the following variables,
\begin{gather*}
x=z^{-2}\qquad\text{and}\qquad y=z^{-2m-1}\prod_{i=1}^m\big(1-(\kappa_iz)^2\big),
\end{gather*}
which gives the relation defining a \emph{singular} hyperelliptic curve,
\begin{gather*}
\mathcal{C}=\bigg\{(x,y)\in \C^2\colon {\mathcal{F}(x,y)}:=y^2-x\prod_{i=1}^m\big(x-\kappa_i^2\big)^2=0\bigg\}.
\end{gather*}
Then the set $\mathcal{A}$ in the variables $(x,y)$ gives a coordinate ring of the curve,
\begin{gather*}
\RR=\C[x,y]/(\mathcal{F}(x,y))={\rm Span}_\C\{{\mathcal{A}}\}.
\end{gather*}
A basis of $\RR$ can be given by the following set of graded elements,
\begin{center} \renewcommand{\arraystretch}{1.2}
\begin{tabular}{c| c| l| c| c| c| c| c| c| c| c }
\hline
0&1&2 &3&$4$ &$\cdots$ &$2m$& $2m+1$ &$2m+2$&$2m+3$&$\cdots$ \\
 \hline
1&$*$&$x$ &$*$&$x^2$&$\cdots$ &$x^m$&$y$ & $x^{m+1}$ &$xy$&$\cdots$\\
\hline
\end{tabular}
\end{center}
which gives the numerical semigroup of type $\langle 2,2m+1\rangle$.

Thus, the soliton solutions of the $2$-reduction (KdV reduction) correspond to the ``double'' point degeneration of the \emph{smooth} hyperelliptc curve,
\begin{gather*}
\tilde{\mathcal{C}}=\bigg\{(x,y)\in\C^2\colon\tilde{\mathcal{F}}(x,y):=
y^2-x\prod_{j=1}^{2m}\big(x-\beta_{j}^2\big)=0\bigg \}.
\end{gather*}
The degenerated curve $\mathcal{C}$ is then given by the double point limit,
\begin{gather*}
\beta_j^2,\beta_{m+j}^2\longrightarrow \kappa_j^2\qquad\text{for all}\quad j=1,2,\ldots,m.
\end{gather*}
This is a well-known result showing that each KdV soliton solution is obtained by a double point limit of
the theta function associated with the hyperelliptic curve~\cite{Mu:84, SW:85}. Also note that we~have
\begin{gather*}
H^0\big(\tilde{\mathcal{C}}, \mathcal{O}(*\infty)\big)= \C[x,y]/\big(\tilde{\mathcal{F}}(x,y)\big).
\end{gather*}
We remark here that the type of the numerical semigroup remains the same as that of the singular curve.

We have the following theorem of the Schur expansion of the $\tau$-function:
\begin{Theorem}\label{thm:2reductionSchur}
The $\tau$-function associated with the numerical $($hyperelliptic$)$ semigroup $S=\langle 2,2m+1\rangle$ has the expansion
\begin{gather*}
\tau(t)=S_{\lambda}(t)+\sum_{\mu\supset \lambda}c_{\hat\mu}(\kappa)S_{\mu}(t),
\end{gather*}
where $\lambda=(m,m-1,\ldots,1)$ and each $\mu$ is obtained by adding a diagram
consisting of several $\young[2][4]$ tiles to $\lambda$. The coefficients $c_{\hat\mu}(\kappa)$
are the Schur polynomials associated to the Young diagrams $\hat\mu$ determined by the structure of the added diagram $\mu$ in terms of $\young[2][4]$ tiles, that is, $\hat\mu$ is obtained by adding $\young[1][4]$ to the previous diagram in the same manner as in the diagram $\mu$
$($see the example below$)$.
\end{Theorem}
\begin{proof}
 First, we note that the coefficients $c_{\hat\mu}(\kappa)$ with $\hat\mu=(\hat\mu_1,\hat\mu_2,\ldots,\hat\mu_m)$ is given by the Schur polynomial,
\begin{gather*}
c_{\hat\mu}(\kappa)=\left|\begin{matrix}
h_{\hat\mu_1}(\kappa) &h_{\hat\mu_2+1}(\kappa)&\cdots &h_{\hat\mu_m+m-1}(\kappa)\\
h_{\hat\mu_1-1}(\kappa)&h_{\hat\mu_2}(\kappa)&\cdots & h_{\hat\mu_m+m-2}(\kappa)\\
\vdots &\vdots &\ddots & \vdots\\
h_{\hat\mu_1-m+1}(\kappa)&h_{\hat\mu_2-m+2}(\kappa)&\cdots&h_{\hat\mu_m}(\kappa)
\end{matrix}\right|,
\end{gather*}
where $h_\alpha(\kappa)$ is the complete homogeneous symmetric polynomials of $\big\{\kappa_1^2,\kappa_2^2,\ldots,\kappa_m^2\big\}$, which can be shown by Lemma~\ref{lem:hook} below (equivalent to the Giambelli's formula)
Then Theorem~\ref{thm:2reductionSchur} can be proven by using the Binet--Cauchy formula for
the $\tau$-function~\eqref{eq:tauP} and the Giambelli identity for the Schur polynomial in the Frobenius notation,
\begin{gather*}
c_{\hat\mu}=c_{(a_1,a_2,\ldots,a_n|b_1,b_2,\ldots,b_n)}=\det \big(c_{(a_i|b_j)}\big),
\end{gather*}
where $\hat\mu=(\hat\mu_1,\hat\mu_2,\ldots,\hat\mu_m)$ is uniquely determined by the Frobenius notation
$(a_1,\ldots,a_n|b_1$, $\ldots,b_n)$, e.g.,\ $\hat\mu_i=a_i+i$ for $1\le i\le n\le m=b_1+1$
(see~\cite{Mac:95}).
\end{proof}

\begin{Lemma}\label{lem:hook}
The reduced row echelon form of the $m\times \infty$ Vandermonde matrix $V=\big(x_i^{j-1}\big)$ for~$1\le i\le m$ and
$1\le j$ is given by
\begin{gather*}
V_0^{-1}V=\begin{pmatrix}
1&0& \ldots &0&(-1)^{m-1} c_{(0|m-1)} & (-1)^{m-1}c_{(1|m-1)}& (-1)^{m-1}c_{(2|m-1)}&\cdots\\
0&1&\cdots& 0&(-1)^{m-2}c_{(0|m-2)}& (-1)^{m-2}c_{(1|m-2)}& (-1)^{m-2}c_{(2|m-2)}&\cdots\\
\vdots&\vdots&\ddots&\vdots &\vdots &\vdots &\vdots&\vdots \\
0& 0& \cdots & 1 & c_{(0|0)} & c_{(1|0)} &c_{(2|0)} & \cdots
\end{pmatrix}\!,
\end{gather*}
where $V_0=\big(x_i^{j-1}\big)_{1\le i,j\le m}$ and $c_{(a|b)}(x)$ is the Schur polynomial of $x=(x_1,\ldots,x_m)$ associated with the hook Young diagram $\lambda$ of $(a|b)$ type, i.e., $\lambda=(1+a,\underbrace{1,\ldots,1}_{b})$, e.g., $(3|2)$ is $\young[4,1,1][4]$.
\end{Lemma}

\begin{proof}
First recall that the Schur polynomial $c_\lambda(x)$ of $(x_1,\ldots,x_m)$ can be also defined by
Jacobi's bialternant formula,
\begin{gather*}
c_\lambda(x)=\frac{a_{\lambda+\delta}(x)}{a_\delta(x)}\qquad\text{with}\quad
a_{\delta}(x)=|V_0|\quad\text{and}\quad a_{\lambda+\delta}(x)=\det \big(x_i^{\lambda_{m-j+1}+j-1}\big),
\end{gather*}
where $\lambda=(\lambda_1,\ldots,\lambda_m)$. The $(r,s)$-element of the row reduced echelon form $V_0^{-1}V$ for $1\le r\le m$ and $s>m$ is given by
\begin{gather*}
\big(V_0^{-1}V\big)_{r,s}=\frac{1}{a_\delta(x)}\left| \begin{matrix}
1 & x_1 &\cdots &x_1^{r-2}&x_1^{s-1}&x_1^r&\cdots & x_1^{m-1}\\
1 & x_2 &\cdots &x_2^{r-2}&x_2^{s-1}&x_2^r&\cdots & x_2^{m-1}\\
\vdots & &\vdots & \vdots&\vdots&\vdots& &\vdots\\
1 & x_m &\cdots &x_m^{r-2}&x_m^{s-1}&x_m^r&\cdots & x_m^{m-1}
\end{matrix}\right|,
\end{gather*}
which is expressed by the Schur polynomial $c_{(s-m-1|m-r)}(x)$. That is, we have
\begin{gather*}
\big(V_0^{-1}V\big)_{r,s}=(-1)^{m-r}c_{(s-m-1|m-r)}(x),
\end{gather*}
which is the desired formula.
\end{proof}

\begin{Example} \label{ex:2red} Consider the case {$S=\langle2, 7\rangle$, $(m=3)$}.
The $\tau$-function associated with $S$ is given~by
\begin{gather*}
\tau(t)=S_{\young[3,2,1][3]}(t)+c_{\young[1][3]}(\kappa)S_{\young[5,2,1][3]}(t)
+c_{\young[2][3]}(\kappa)S_{\young[7,2,1][3]}(t)+c_{\young[1,1][3]}(\kappa)S_{\young[5,4,1][3]}(t)
\\ \hphantom{\tau(t)=}
{}+c_{\young[3][3]}(\kappa)S_{\young[9,2,1][3]}(t) +c_{\young[2,1][3]}(\kappa)S_{\young[7,4,1][3]}(t) +c_{\young[1,1,1][3]}(\kappa)S_{\young[5,4,3][3]}(t)+\cdots,
\end{gather*}
where $c_{\hat\mu}(\kappa)$ is the Schur polynomial of $\big\{\kappa_1^2,\kappa_2^2,\kappa_3^2\big\}$
with the Young diagram $\hat\mu=(\hat\mu_1,\hat\mu_2,\hat\mu_3)$. Note here that the Young diagram $\mu$ in $S_\mu$ increases by $\young[2][4]$, while $\hat\mu$ in $c_{\hat\mu}$ increases by $\young[1][4]$
in~the same way.
\end{Example}

\begin{Remark}
For the soliton solution for $(l,m,l-1)$ (i.e., $k=l-1$), the coefficients $c_{\hat\mu}(\kappa)$ are
given by the Schur polynomials of $\big(\kappa_1^l,\ldots,\kappa_m^l\big)$ associated with the Young diagrams $\hat\mu$. However,
for the generalized soliton solutions, $c_{\hat{\mu}}(\kappa)$ are not the Schur polynomials in general
(see the example below).
\end{Remark}

\subsection{3-reductions: Boussinesq reduction}
We consider the numerical semigroups
of type $\langle 3, 3m+1\rangle$ and $\langle 3, 3m+1, 3m+2\rangle$.

\subsubsection[S=<3, 3m+1, 3m+2>]{$\boldsymbol{S=\langle 3, 3m+1, 3m+2\rangle}$}

We have $l=3$ and $k=2$, which gives $n_{l,k}=1$. The Frobenius number $F(S)$ and the genus $g(S)$ are
\begin{gather*}
F(s)=3m-1\qquad\text{and}\qquad g(S)=2m.
\end{gather*}
The Young diagram is then given by
\begin{center}
\young[6,4,2][6]
\end{center}
for the case with $m=3$. The soliton solution here has the following $\kappa$-parameters and
the $m\times 3m$ matrix~$A$,
\begin{gather*}
\big(\kappa_1\Omega_3^1,\kappa_2\Omega_3^1,\ldots,\kappa_m\Omega_3^1\big)\qquad\text{and}\qquad A=I_m\otimes\Omega_3^1,
\end{gather*}
where $\Omega_3^k:=\big(1,\omega_3^k,\omega_3^{2k}\big)$ with $\omega_3=\exp\frac{2\pi {\rm i}}{3}$.
The matrix $K$ is an $\infty \times 3m$ given by
\begin{gather*}
K=\begin{pmatrix}
1_3 & 1_3 & \cdots & 1_3\\
\kappa_1\Omega_3^1&\kappa_2\Omega_3^1&\cdots &\kappa_m\Omega_3^1\\[.5ex]
\kappa_1^2\Omega_3^2 &\kappa_2^2\Omega_3^2&\cdots &\kappa_m^2\Omega_3^2\\[.5ex]
\kappa_1^31_3&\kappa_2^31_3&\cdots &\kappa_m^31_3\\
\vdots & \vdots&\vdots &\vdots
\end{pmatrix}\!,
\end{gather*}
where $1_3=\Omega_3^0=(1,1,1)$.
Then using the orthogonality $\Omega_3^i\cdot\Omega_3^j=3\delta_{i+j,3}$ and column operations, the matrix $KA^{\rm T}$ can be expressed as
\begin{gather}\label{eq:3reductionK}
KA^{\rm T}\equiv\begin{pmatrix}
0 & 0 & \cdots& 0\\
0 & 0 & \cdots & 0\\
1 & 1 & \cdots & 1\\
0 & 0 & \cdots & 0\\
0 & 0 & \cdots & 0\\
\kappa_1^3&\kappa_2^3&\cdots &\kappa_m^3\\
\vdots &\vdots &\vdots&\vdots
\end{pmatrix}\!.
\end{gather}
The embedding $\sigma_K$ then gives an element $U={\rm Span}_\C\{\mathcal{B}\}\in$UGM with
\begin{gather*}
\mathcal{B}=\bigg\{\phi_{-j+1}(z) =\frac{z^{3-m}}{1-(\kappa_jz)^3}~(\text{for}~j=1,\ldots,m),\ \phi_{-i}(z)=z^{-i}~(\text{for}~i\ge m)\bigg\}.
\end{gather*}
Multiplying $\mathcal{B}$ by $z^{-g}\prod_{j=1}^m\big(1-(\kappa_jz)^3\big)$ with $g=g(S)=2m$, we find a basis of the subspace of meromorphic functions having pole only at $z=0$,
\begin{gather*}
\mathcal{A}=\bigg\{ 1,z^{-3}, z^{-6},\ldots, z^{3-3m}, z^{-3m}, z^{-3m-i}\prod_{j=1}^m\big(1-(\kappa_jz)^3\big)\ (\text{for}~i\ge 1)\bigg\}.
\end{gather*}
Now we set
\begin{gather*}
x=z^{-3},\qquad
\text{and}\qquad
y_1=z^{-3m-1}\prod_{j=1}^m\big(1-(\kappa_jz)^3\big),\qquad
y_2=z^{-3m-2}\prod_{j=1}^m\big(1-(\kappa_jz)^3\big).
\end{gather*}
These variables then satisfy the following polynomial relations for $(x,y_1,y_2)$,
\begin{gather}\label{eq:relations}
y_1^2=y_2F(x),\qquad
y_1y_2=F(x)G(x),\qquad
y_2^2=y_1G(x),
\end{gather}
where $F(x)=\prod_{j=1}^m\big(x-\kappa_j^3\big)$ and $G(x)=xF(x)$. These relations are the minors
of the following $2\times 3$ matrix~\cite{KMP:13, P:74},
\begin{gather*}
\begin{pmatrix}
y_1 & y_2 & G(x)\\
F(x) & y_1 & y_2
\end{pmatrix}\!,
\end{gather*}
(see~\eqref{eq:2xl}). Since the affine algebraic set defined by the functions in~\eqref{eq:relations} can be shown to be irreducible, we have
the coordinate ring,
\begin{gather}\label{eq:3ring}
\RR=\C[x,y_1,y_2]/\mathcal{P},
\end{gather}
where $\mathcal{P}$ is the prime ideal defined in~\eqref{eq:relations}~\cite{KMP:13}. Then one should note that
the ring defines a~singular space curve whose affine part is given by $\text{Spec}(\RR)$.
The ring~\eqref{eq:3ring} gives a subspace of the meromorphic functions on the singular space curves having pole only at $\infty$, i.e., $\RR={\rm Span}_\C\big\{{\tilde{\mathcal{A}}}\big\}$ with
\begin{gather*}
\tilde{\mathcal{A}}=\big\{1, x, x^2, \ldots, x^{m}, y_1, y_2, x^{m+1}, xy_1, xy_2, x^{m+2}, x^2y_1, x^2y_2,\ldots\big\}.
\end{gather*}
In Section~\ref{sec:deformation}, we will discuss a deformation of this singular space curve
by setting
\begin{gather*}
F(x)\longrightarrow \tilde F(x)=\prod_{j=1}^m\big(x-\beta_j^3\big),\qquad
G(x)\longrightarrow \tilde G(x)=
x\prod_{j=1}^m\big(x-\beta_{m+j}^3\big),
\end{gather*}
where $\{\beta_j^3\colon j=1,\ldots,2m\}$ are distinct constants. The polynomial ring $\tilde{\RR}$
given by $\C[x,y_1,y_2]/\tilde{\mathcal{P}}$ with
\begin{gather*}
\tilde{\mathcal{P}}=\big\{y_1^2=y_2\tilde F(x),\,y_1y_2=\tilde F(x)\tilde G(x),\,y_2^2=y_1\tilde G(x)\big\},
\end{gather*}
defines a smooth space curve $\tilde{\mathcal{C}}$ (see Section~\ref{sec:deformation}). We~also note
\begin{gather*}
\tilde\RR=\C[x,y_1,y_2]/\tilde{\mathcal{P}}=H^0\big(\tilde{\mathcal{C}},\mathcal{O}(*\infty)\big).
\end{gather*}
Thus each soliton solution considered here is associated with a coalescence of two moduli para\-me\-ters
of the smooth curve $\tilde{\mathcal{C}}$, i.e., $\beta_j^3,\beta_{m+j}^3\to\kappa_j^3$. We~remark that the singularity obtained by~the coalescence is ``not'' an ordinary double point, but an \emph{ ordinary 3-tuple point singularity} as~in~\cite{BG:80}.

Similar to Theorem~\ref{thm:2reductionSchur}, the Schur expansion of the $\tau$-function can be obtained using the matrix $KA^{\rm T}$ in~\eqref{eq:3reductionK}. For example, consider $S=\langle 3, 7, 8\rangle$, ($m=2$). We~have
\begin{gather*}
\tau(t)=S_{\young[4,2][3]}(t)+c_{\young[1][3]}(\kappa)S_{\young[7,2][3]}(t)
+c_{\young[1,1][3]}(\kappa)S_{\young[7,5][3]}(t)+c_{\young[2][3]}(\kappa)S_{\young[10,2][3]}(t)
\\ \hphantom{\tau(t)=}
{}+ c_{\young[2,1][3]}(\kappa)S_{\young[10,5][3]}(t)+c_{\young[2,2][3]}(\kappa)
S_{\young[10,8][3]}(t)+\cdots,
\end{gather*}
where $c_{\mu}(\kappa)$ is the Schur polynomial of $\big(\kappa_1^3,\kappa_2^3\big)$ associated to
the Young diagram $\mu$. Here the Young diagram $\mu$ in $S_\mu$ increases by adding
$\young[3][4]$ to the previous one, and the diagram $\hat\mu$ in $c_{\hat\mu}$ increases by adding
$\young[1][4]$ to the previous one (cf.\ Example~\ref{ex:2red}).

\subsubsection[S=<3,3m+1>]{$\boldsymbol{S=\langle 3,3m+1\rangle}$}
Here we consider the case with $m=2$, i.e., $S=\langle 3,7\rangle$ (the general case should be obvious). In~this case, we have
\begin{gather*}
n_{l,k}=2,\qquad
F(S)=11,\qquad
g(S)=6, \qquad
M(S)=12,\qquad
N(S)=6.
\end{gather*}
The corresponding Young diagram is $\lambda(S)=(6,4,2,2,1,1)$, i.e.,
\begin{center}
\young[6,4,2,2,1,1][6]
\end{center}
The matrix~$A$ is the $6\times 12$ matrix,
\begin{gather*}
A=\begin{pmatrix}
I_2\otimes[\Omega_3]^1 &0
\\
0&I_2\otimes[\Omega_3]^2
\end{pmatrix}\!.
\end{gather*}
The base function~\eqref{eq:base} is given by
\begin{gather*}
\bigg(\bigg[\frac{\partial}{\partial \kappa}\kappa \mathsf{E}(\kappa_1)\bigg]_2,
\bigg[\frac{\partial}{\partial \kappa}\kappa \mathsf{E}(\kappa_2)\bigg]_2,
[\mathsf{E}(\kappa_1)]_2,[\mathsf{E}(\kappa_2)]_2\bigg),
\end{gather*}
where $\mathsf{E}(\kappa_i)=(E_1(t,\kappa_i),\ldots,E_l(t,\kappa_i))$ with
$E_j(t,\kappa_i)=\exp\big(\sum_{n=1}^\infty \big(\kappa_i\omega_3^{j-1}\big)^nt_n\big)$.
Then the Vandermonde matrix $K$ is the $\infty\times 12$ matrix,
\begin{gather*}
K=\begin{pmatrix}
1_3 & 1_3&1_3&1_3\\
2\kappa_1\Omega_3^1&2\kappa_2\Omega_3^1&\kappa_1\Omega_3^1&\kappa_2\Omega_3^1\\[.5ex]
3\kappa_1^2\Omega_3^2&3\kappa_2^2\Omega_3^2&\kappa_1^2\Omega_3^2&\kappa_2^2\Omega_3^2\\[.5ex]
4\kappa_1^3\Omega_3^3&4\kappa_2^3\Omega_3^3&\kappa_1^3\Omega_3^3&\kappa_2^3\Omega_3^3\\
\vdots &\vdots&\vdots&\vdots
\end{pmatrix}\!.
\end{gather*}
The $\infty\times 6$ matrix $KA^{\rm T}$ is then given by, after column operations,
\begin{gather*}
KA^{\rm T} \equiv \left(\begin{array}{llllllllllllll}
0 & 1 & 0 & 0 & \kappa_1^3 & 0 & 0 &\kappa_1^6&0 & 0 &\kappa_1^9 & 0&0& \cdots\\[0.5ex]
0 & 1 & 0 & 0 & \kappa_2^3 & 0 & 0 &\kappa_2^6&0 & 0 &\kappa_2^9 & 0&0& \cdots\\[0.5ex]
0 & 0 & 1 & 0 & 0 & \kappa_1^3 & 0 & 0 & \kappa_1^6& 0 & 0 & \kappa_1^9 &0& \cdots \\[0.5ex]
0 & 0 & 1 & 0 & 0 & \kappa_2^3 & 0 & 0 & \kappa_2^6& 0 & 0 & \kappa_2^9 &0& \cdots \\[0.5ex]
0 & 0 & 0 & 0 & 0 & 1 & 0 & 0 & 2\kappa_1^3 & 0 & 0 &3\kappa_1^6 &0& \cdots \\[0.5ex]
0 & 0 & 0 & 0 & 0 & 1 & 0 & 0 & 2\kappa_2^3 & 0 & 0 &3\kappa_2^6 &0& \cdots
\end{array}\right)^{\rm T}.
\end{gather*}
Then the Sato frame corresponding to the generalized soliton solution is given by
\begin{gather*}
\mathcal{B}=\bigg\{
\frac{1}{(1-(\kappa_jz)^3)^2},\frac{z^{-3}}{1-(\kappa_jz)^3}, \frac{z^{-4}}{1-(\kappa_jz)^3},~(j=1,2)\ \text{and}\ z^{-i}~(\text{for}~i\ge 6)\bigg\}.
\end{gather*}
Multiplying $\mathcal{B}$ by $z^{-6}F(z)^2$ with $F(z)=\prod_{j=1}^2\big(1-(\kappa_jz)^3\big)$, we have
\begin{gather*}
\mathcal{A}=\big\{1,z^{-3},z^{-6},z^{-7}F(z),z^{-9},z^{-10}F(z),z^{-i-6}F(z)~(\text{for}~i\ge 6)\big\}\subset\C\big[z^{-1}\big].
\end{gather*}
We then set
\begin{gather*}
x=z^{-3},\qquad
\text{and}\qquad
y=z^{-7}F(z)=z^{-1}\prod_{j=1}^2\big(x-\kappa_j^3\big),
\end{gather*}
which defines a singular (plane) trigonal curve,
\begin{gather*}
\mathcal{C}=\bigg\{(x,y)\in\C^2\colon y^3=x\prod_{j=1}^2\big(x-\kappa_j^3\big)^3\bigg\}.
\end{gather*}
That is, we have a coordinate ring,
\begin{gather*}\RR=\C[x,y]/\mathcal{C}={\rm Span}_{\C}\big\{\tilde{\mathcal{A}}\big\},
\end{gather*}
where the basis $\tilde{\mathcal{A}}$ of a set of meromorphic functions is given by
\begin{gather*}
\tilde{\mathcal{A}}=\big\{1,x,x^2,y,x^3,xy,x^4,x^2y,x^3y,y^2,\ldots\big\}.
\end{gather*}
Note that the singular curve can be obtained by taking a triple point degeneration of
a trigonal curve (called $(3,4)$-curve),
\begin{gather*}
y_1^3=x\prod_{i=1}^{3}\big(x-\beta_i^3\big)\big(x-\beta_{3+i}^3\big)\qquad\text{with}\quad \beta_{3j-2}^3,\,\beta_{3j-1}^3,\,\beta_{3j}^3\longrightarrow \kappa_j^3\quad(j=1,2).
\end{gather*}
Thus, the plain curve for the generalized soliton solution associated with $S=\langle 3,7\rangle$ has two ordinary triple point singularities.
This case was recently studied in~\cite{Na:18a}.

The Schur expansion of the $\tau$-function in this case is given by
\begin{gather*}
\tau(t)=S_{\young[6,4,2,2,1,1][3]}(t)+2c_{\young[1][3]}(\kappa)S_{\young[9,4,2,2,1,1][3]}(t) -c_{\young[1][3]}(\kappa)S_{\young[6,4,4,3,1,1][3]}(t)+c_{\young[2][3]}(\kappa)S_{\young[6,6,5,3,1,1][3]}(t)
\\[1ex] \hphantom{\tau(t)=}
{}-2(c_{\young[2][3]}+c_{\young[1,1][3]})(\kappa)S_{\young[9,4,4,3,1,1][3]}(t)+(3c_{\young[2][3]} +c_{\young[1,1][3]})(\kappa)S_{\young[12,4,2,2,1,1][3]}(t)
\\[1ex] \hphantom{\tau(t)=}
{}+(c_{\young[2][3]}+3c_{\young[1,1][3]})(\kappa)S_{\young[9,7,2,2,1,1][3]}(t)+\cdots,
\end{gather*}
where each $c_{\mu}(\kappa)$ is the Schur polynomial of $\big(\kappa_1^3,\kappa_2^3\big)$ associated with the Young diagram~$\mu$.
 Note that the Young diagrams $\lambda$ in $S_{\lambda}(t)$
are obtained by adding $\young[3][4]$ or $\young[2,1][4]$ to the previous diagrams.
Also note that each coefficient is a homogeneous polynomial but not a single Schur polynomial.
The general formula of the coefficients $c_{\hat\mu}(\kappa)$ for $\kappa=(\kappa_1,\ldots,\kappa_m)$ will be discussed elsewhere.

\section[Proof of Theorem~5.1]{Proof of Theorem~\ref{thm:Main}}\label{sec:Proof}

We first summarize the notations and then give several steps to prove the theorem.

(1) The $lmq$-dimensional base for the exponential functions is given by
\begin{gather*}
\hat{\mathsf{E}}=\big(\big[\hat{\mathsf{E}}^{(q-1)}\big]_m, \big[\hat{\mathsf{E}}^{(q-2)}\big]_m, \ldots, \big[\hat{\mathsf{E}}^{(0)}\big]_m\big)\qquad
\text{with}\quad q=n_{l,k}=\bigg\lceil\frac{l-1}{k}\bigg\rceil,
\end{gather*}
where $[\hat{\mathsf{E}}^{(j)}]_m$ is the $lm$-dimensional row vector defined by
\begin{gather*}
[\hat{\mathsf{E}}^{(j)}]_m=\big(\hat{\mathsf{E}}^{(j)}(\kappa_1), \hat{\mathsf{E}}^{(j)}(\kappa_2), \ldots, \hat{\mathsf{E}}^{(j)}(\kappa_m)\big)\qquad
\text{with}\quad
\hat{\mathsf{E}}^{(j)}(\kappa_i)=\frac{1}{j!}\frac{\partial^j}{\partial \kappa_i^j}\kappa_i^j\mathsf{E}(\kappa_i).
\end{gather*}
Each $l$-dimensional vector $\mathsf{E}(\kappa_i)$ is given by
\begin{gather*}
\mathsf{E}(\kappa_i)=(E_1(t,\kappa_i),E_2(t,\kappa_i),\ldots,E_l(t,\kappa_i))\qquad
\text{with}\quad
E_p(t,\kappa_i)=\exp\bigg(\sum_{n=1}^\infty \big(\kappa_i\omega_l^{p-1}\big)^nt_n\bigg).
\end{gather*}
The Schur expansion of $\hat{\mathsf{E}}^{(j)}(\kappa_i)$ is expressed by
\begin{gather*}
\hat{\mathsf{E}}^{(j)}(\kappa_i)=(1,p_1(t),p_2(t),p_3(t),\ldots)\hat{K}^{(j)}(\kappa_i),
\end{gather*}
where $\hat{K}^{(j)}$ is the $\infty\times l$ matrix whose $n$th row is given by
\begin{gather*}
\text{Row}_n\big(\hat{K}^{(j)}(\kappa_i)\big)=\binom{n+j}{j}\kappa_i^n\Omega_l^n
\qquad
\text{for}\quad
n=0,1,2,\ldots.
\end{gather*}
Here $\Omega_l^n=\big(1,\omega_l^n,\omega_l^{2n},\ldots,\omega_l^{n(l-1)}\big)$ with $\omega_l=\exp(2\pi {\rm i}/l)$. The Schur expansion of the base $\hat{\mathsf{E}}$ is then expressed by
\begin{gather*}
\hat{\mathsf{E}}=(1,p_1(t),p_2(t),\ldots)\hat{K},
\end{gather*}
where
\begin{gather*}
\hat{K}=\big(\big[\hat{K}^{(q-1)}\big]_m,\ldots,\big[\hat{K}^{(0)}\big]_m\big)\qquad
\text{with}\quad
[\hat{K}^{(j)}]_m=\big(\hat{K}^{(j)}(\kappa_1),\ldots,\hat{K}^{(j)}(\kappa_m)\big).
\end{gather*}

(2) The $A$-matrix is the $N\times mlq$ matrix given by
\begin{gather*}
A=\begin{pmatrix}
I_m\otimes [\Omega_l]^{n_1} & 0&\cdots & 0\\
0 & I_m\otimes [\Omega_l]^{n_2}&\cdots &0\\
\vdots &\ddots &\ddots &\vdots\\
0 &\cdots &0&I_m\otimes [\Omega_l]^{n_q}
\end{pmatrix}
\quad\text{with}\quad [\Omega_l]^n=\begin{pmatrix}
\Omega_l^1 \\ \Omega_l^2 \\ \vdots \\ \Omega_l^n
\end{pmatrix}\!,
\end{gather*}
where $N=\sum_{i=1}^qmn_i$ with $n_i=k(i-1)+1$.
Calculating $\hat{K}A^{\rm T}$, we have
\begin{gather*}
\hat{K}A^{\rm T}=\big(\big[B^{(q-1)}\big]^{n_1}_m,\big[B^{(q-2)}\big]^{n_2}_m,\ldots,\big[B^{(0)}\big]^{n_q}_m\big).
\end{gather*}
Each element $[B^{(j)}]^{n_{q-j}}_m$ is the $\infty\times mn_{q-j}$ matrix,
\begin{gather*}
\big[B^{(j)}\big]^{n_{q-j}}_m=\big(\big[B^{(j)}(\kappa_1)\big]^{n_{q-j}},\big[B^{(j)}(\kappa_2)\big]^{n_{q-j}},\ldots, \big[B^{(j)}(\kappa_m)\big]^{n_{q-j}}\big).
\end{gather*}
The $n$th column of the $\infty\times n_{q-j}$ matrix $\big[B^{(j)}(\kappa_i)\big]^{n_{q-j}}$ is given by
\begin{gather*}
\text{Col}_n\big(\big[B^{(j)}(\kappa_i)\big]^{n_{q-j}}\big)
=\bigg(\!\underbrace{0,\ldots,0}_{l-n},l\binom{l+j-n}{j}\kappa_i^{l-n},\underbrace{0,\ldots,0}_{l-1}, l\binom{2l+j-n}{j}\kappa_i^{2l-n},0,\ldots \!\bigg)^{\rm T}\!.
\end{gather*}

(3) Consider the embedding $\sigma_K\colon A\mapsto \widetilde{\hat{K}A^{\rm T}}$. Then the corresponding element of UGM can be expressed as follows: For each $0\le j\le q-1$ and $1\le n\le n_{q-j}$, we have
\begin{gather*}
\psi_{j,n}(z,\kappa_i):=\big(z^{-N+1},z^{-N+2},\ldots,z^{-1},1,z^1,\ldots\big) \text{Col}_n\big(\big[B^{(j)}(\kappa_i)\big]^{n_{q-j}}\big)
\\ \hphantom{\psi_{j,n}(z,\kappa_i)}
{}=lz^{-N+1}\sum_{r=1}^\infty \binom{rl+j-n}{j}(\kappa_iz)^{rl-n}
=lz^{-N+1}\frac{1}{j!}\frac{{\rm d}^j}{{\rm d}u^j} \bigg(\sum_{r=1}^{\infty}u^{rl+j-n}\bigg)\bigg|_{u=\kappa_iz}
\\ \hphantom{\psi_{j,n}(z,\kappa_i)}
{}= lz^{-N+1}\frac{1}{j!}\frac{{\rm d}^j}{{\rm d}u^j}\bigg(\frac{u^{l+j-n}}{1-u^l}\bigg)\bigg|_{u=\kappa_iz},
\end{gather*}
which form the set of $N$ elements in the base $\{\phi_{-i}(z)\colon i\in\N_0\}$ for a point of UGM, i.e., we assign
\begin{gather*}
\{\phi_{-i}(z)\colon 0\le i\le N-1\}=\{\psi_{j,n}(z,\kappa_i)\colon 0\le j\le q-1,\,1\le n\le n_{q-j},\, 1\le i\le m\}.
\end{gather*}
And the rest in the base $\{\phi_{-i}(z)\colon i\in \N_0\}$ is given by $\{\phi_{-i}(z)=z^{-i}\colon i\ge N\}$.

(4) Then we note that the set $\{\phi_{-i}(z)\colon 0\le i\le N-1\}$ as a base of an $N$-dimensional vector space is equivalent to
\begin{gather*}
\bigg\{\frac{z^{-N-s_j+l}}{(1-(\kappa_iz)^l)^{j+1}}\colon 0\le j< q,\,
0\le s_j< n_{q-j},\, 1\le i\le m\bigg\}.
\end{gather*}
The other elements of the base are given by $\{\phi_{-i}(z)=z^{-i}\colon i\ge N\}$.

Now multiplying the base by $z^{N-mql}\prod_{i=1}^m\big(1-(\kappa_iz)^l\big)^q$, we have a set of polynomials in $z^{-1}$ of the form $\{\phi_{-i}(z)=z^{-i}F_i(z)\colon i\in\N_0\}$, where $F_i(z)$ is a polynomial in $z$ with $F_0(z)=1$ and $\deg (F_i(z))<i$ for $i\ge 1$. More precisely, we have the following:
First we have
\begin{gather*}
\frac{z^{-(mq-1)l-s_j}}{\big(1-(\kappa_iz)^l\big)^{j+1}}\prod_{p=1}^m\big(1-(\kappa_pz)^l\big)^q
=z^{-(mq-1)l-s_j}\prod_{p\ne i}\big(1-(\kappa_pz)^l\big)^q \big(1-(\kappa_iz)^l\big)^{q-1-j}.
\end{gather*}
Let us now express a set of basis for these meromorphic sections.
\begin{itemize}\itemsep=0pt
\item[$(i)$] For $s_j=0$, we have $0\le j\le q-1$, and one can have an equivalent base for this case, that~is,
\begin{gather*}
\bigg\{z^{-(mq-1)l}\prod_{p\ne i}\big(1-(\kappa_pz)^l\big)^q \big(1-(\kappa_iz)^l\big)^{q-1-j}\colon 0\le j\le q-1,\, 1\le i\le m\bigg\}
\\ \qquad
{}\equiv\big\{1,z^{-l},z^{-2l},\ldots,z^{-(mq-1)l}\big\}=\big\{z^{-pl}\colon 0\le p\le mq-1\big\}.
\end{gather*}
Here the number of elements is $mq$.
\item[$(ii)$] For each $s_j$ with $1\le s_j\le k$, we have $0\le j\le q-2$. Then we write
\begin{gather*}
z^{-(mq-1)l-s_j}\prod_{p\ne i}\big(1-(\kappa_pz)^l\big)^q \big(1-(\kappa_iz)^l\big)^{q-1-j}
\\ \qquad
{}=z^{-(mq-1)l-s_j}\prod_{i=1}^m\big(1-(\kappa_iz)^l\big)\bigg(\prod_{p\ne i}\big(1-(\kappa_pz)^l\big)^{q-1} \big(1-(\kappa_iz)^l\big)^{q-2-j}\bigg),
\end{gather*}
and an equivalent base can be expressed as
\begin{gather*}
\bigg\{z^{-(mq-1)l-s_j}\prod_{i=1}^m\big(1-(\kappa_iz)^l\big)\bigg(\prod_{p\ne i}\big(1-(\kappa_pz)^l\big)^{q-1} \big(1-(\kappa_iz)^l\big)^{q-2-j}\bigg)\bigg\}
\\ \qquad
{}\equiv\bigg\{\prod_{i=1}^m\big(1-(\kappa_iz)^l\big) z^{-pl-s_j}\colon m\le p\le mq-1,\, 1\le i\le m\bigg\}.
\end{gather*}
Here the number of elements in the base is $m(q-1)$ for each $1\le s_j\le k$.
\item[$(iii)$] For each $s_j$ with $k+1\le s_j\le 2k$, we have $0\le j\le q-3$. Then
an equivalent base is
\begin{gather*}
\bigg\{z^{-(mq-1)l-s_j}\prod_{i=1}^m\big(1-(\kappa_iz)^l\big)^2\bigg(\prod_{p\ne i}\big(1-(\kappa_pz)^l\big)^{q-2} \big(1-(\kappa_iz)^l\big)^{q-3-j}\bigg)\bigg\}
\\ \qquad
{}\equiv\bigg\{\prod_{i=1}^m\big(1-(\kappa_iz)^l\big)^2z^{-pl-s_j}\colon 2m\le p\le mq-1,\, 1\le i\le m\bigg\}.
\end{gather*}
This continues to $(q-2)k+1\le s_j\le (q-1)k$ for just $j=0$, and this gives
\begin{gather*}
\bigg\{z^{-(mq-1)l-s_j}\prod_{i=1}^m\big(1-(\kappa_iz)^l\big)^{q-1}\prod_{p\ne i}\big(1-(\kappa_pz)^l\big)\bigg\}
\\ \qquad
{}\equiv\bigg\{\prod_{i=1}^m\big(1-(\kappa_iz)^l\big)^{q-1}z^{-pl-s_j}\colon (q-1)m\le p\le mq-1\bigg\},
\end{gather*}
whose number of elements is just $m$ for each $s_0$ in $(q-2)k+1\le s_0\le (q-1)k$.
\end{itemize}

Notice that the total number of elements in the bases in (i) through (iii) is just $N=mq(1+k(q-1)/2)$.
Then we have a base $\mathcal{A}(z)=\mathcal{A}_1(z)\cup\mathcal{A}_2(z)$ for a subspace of $\C\big[z^{-1}\big]$, where
\begin{gather*}
\mathcal{A}_1(z):=\left\{ z^{-p(r)l-s_j(r)}\prod_{p=1}^m\big(1-(\kappa_pz)^l\big)^r\colon
\begin{array}{lllll}
0\le r\le q-1,\\[0.5ex]
rm\le p(r)\le mq-1,\\[0.5ex]
(r-1)k+1\le s_j(r)\le rk
\end{array}\right\}\!,
\\
\mathcal{A}_2(z):=\bigg\{z^{-mql-i+N}\prod_{p=1}^m\big(1-(\kappa_pz)^l\big)^q\colon i\ge N\bigg\}.
\end{gather*}

(5) Consider the index set for the minimal degrees of the polynomials in $\mathcal{A}=\mathcal{A}_1\cup\mathcal{A}_2$, i.e.,
\begin{gather*}
S=\left\{p(r)l+s(r)\colon \begin{array}{lllll}
0\le r\le q-1,\\[0.5ex]
rm\le p(r)\le mq-1,\\[0.5ex]
(r-1)k+1\le s(r)\le rk
\end{array}\right\}\bigcup\,\{mql+i\colon i\ge 0\}.
\end{gather*}
We now show that the set $S$ forms a numerical semigroup of type $\langle l,lm+1,lm+2,\ldots,lm+k\rangle$. It is obvious that $0\in S$ and $|\N_0\setminus S|<\infty$.
The closeness under the addition is given as follows: Write $p(r)$ and $s(r)$ for $1\le r\le q-1$ as
\begin{gather*}
p(r)=rm+\alpha,\qquad 0\le \alpha\le m(q-r)-1,
\\
s(r)=(r-1)k+\beta,\qquad 1\le \beta\le k.
\end{gather*}
Then each element can be expressed as
\begin{gather*}
p(r)l+s(r)=\alpha l+(lm+\beta )+ (r-1)(lm+k),
\end{gather*}
which is in the span $\langle l,lm+1,\ldots,lm+k\rangle$. For the element $mql+i$ with $i\ge 0$, we first write
\begin{gather*}
i=\gamma l+\delta\qquad
\text{with}\quad
0\le \delta\le l-1.
\end{gather*}
Then noting $l-1\le qk$, we have
\begin{gather*}
\delta=\mu k+\nu\qquad
\text{with}\quad 0\le \mu\le q \quad\text{and}\quad 0\le \nu\le k-1,
\end{gather*}
which leads to
\begin{gather*}
mql+i=\mu(ml+k)+\gamma l+m(q-\mu)l+\nu.
\end{gather*}
When $k=1$, we have $\nu=0$ and $mql+i$ is in the span $\langle l,lm+1\rangle$.
For the case with $k>1$ and $\nu>0$, we have $\mu<q$, hence
\begin{gather*}
m(q-\mu)l=m(q-\mu-1)l+ml.
\end{gather*}
Therefore the element $mql+i$ is in the span $\langle l,lm+1,\ldots,lm+k\rangle$ for all $i\ge 0$.

We introduce the following variables,
\begin{gather*}
x=z^{-l}\qquad
\text{and}\qquad
y_i=z^{-ml-i}\prod_{p=1}^m\big(1-(\kappa_pz)^l\big)\qquad
\text{for}\quad
1\le i\le k.
\end{gather*}
Then, the subspace spanned by the basis $\mathcal{A}$ can be expressed by a polynomial ring,
\begin{gather*}
\RR=\C[x,y_1,\ldots,y_k]/\mathcal{P},
\end{gather*}
where the prime ideal $\mathcal{P}$ is encoded in the set of $2\times 2$ minors of the
following $2\times (k+1)$ matrix,
\begin{gather}\label{eq:DegenerateIdeal}
\begin{pmatrix}
y_1 & y_2 &\cdots & y_k & G(x)\\
F(x)^{l-k} & F(x)^{l-k-1}y_1 &\cdots & F(x)^{l-k-1}y_{k-1} & y_1^{l-k-1}y_k
\end{pmatrix}\!.
\end{gather}
Here the functions $F(x)$ and $G(x)$ are given by
\begin{gather*}
F(x)=\prod_{j=1}^m\big(x-\kappa_j^l\big)\qquad\text{and}\qquad G(x)=xF(x).
\end{gather*}
Note that the spectrum ${\rm Spec}(\RR)$ gives the affine part of a singular curve in the $(k+1)$-dimensional space of $(x,y_1,\ldots,y_k)$. We~also remark that this space curve is an irreducible
component of the singular curve defined by
\begin{gather*}
\mathcal{C}=\big\{(x,y_1,\ldots,y_k)\in\C^{k+1}\colon y_i^l=G(x)^iF(x)^{l-i},\, i=1,\ldots,k\big\}.
\end{gather*}

(6) Let~$\lambda(S)$ be the Young diagram for the numerical semigroup $S$.
Using the Binet--Cauchy formula, we can see that the $\tau$-function has the expansion,
\begin{gather*}
\tau(t)=\left|
\begin{pmatrix}
1& p_1&p_2&\cdots&\cdots &\cdots\\
0& 1 & p_1 &p_2&\cdots&\cdots\\
\vdots&\ddots&\ddots &\ddots &\ddots&\vdots\\
0&\cdots& 0&1&p_1&\cdots
\end{pmatrix} \hat K\, A^{\rm T}\right|
=S_\lambda(\hat t)+\sum_{\mu\supset\lambda}c_\mu(\kappa)S_\mu(\hat t),
\end{gather*}
where $\hat t=\{t_n\colon n\ne ml~\text{for any~} m\in\N\}$ and $c_{\mu}(\kappa)$ is a homogeneous symmetric polynomial of~$\big(\kappa_1^l,\ldots,\kappa_m^l\big)$, which is given by the $N\times N$ minor of the matrix $\hat KA^{\rm T}$ with the rows indexed by~the diagram $\mu$. Also note that the Young diagram associated with the matrix~$A$
is just~$\lambda(S)$ (see Theorem~\ref{thm:center}), i.e., $A\in X_{\lambda(S)}\subset {\rm Gr}(N,M)$. The pivot index of $A$ is given by
\begin{gather*}
i_k=s_{k-1}+1\qquad\text{for}\quad 1\le k\le N=mq\bigg(1+\frac{1}{2}k(q-1)\bigg),
\end{gather*}
where $s_j$ for $j=0,1,\ldots,N-1$ are the first $N$ ordered elements in~$S$.

\section{Deformation of the singular space curves for soliton solutions}\label{sec:deformation}
Her we construct a smooth space curve for the numerical semigroup of type
$\langle l,lm+1,\ldots$, $l(m+1)-1\rangle$, i.e., $k=l-1$. Then the smooth curve naturally degenerates to the singular curve associated with a soliton solution of the $l$-th generalized KdV hierarchy.

Let us recall that the singular curve for the soliton solution corresponding to this type has the local coordinates,
\begin{gather*}
x=z^{-l},\qquad
y_i=z^{-i}\prod_{j=1}^m\big(x-\kappa_j^l\big)\qquad
(1\le i\le l-1).
\end{gather*}
Then we have the following algebraic relations (i.e.,~\eqref{eq:DegenerateIdeal} with $k=l-1$),
\begin{gather}\label{eq:prime}
p_{i,j}:=y_iy_{j-1}-y_jy_{i-1}=0\qquad\text{for}\quad 1\le i<j\le l,
\end{gather}
with $y_0:=F(x)=\prod_{j=1}^m(x-\kappa_j^l)$ and $y_l:=G(x)=x F(x)$.
These relations can be expressed by the $2\times 2$ minors of the following $2\times l$ matrix,
\begin{gather*}
\begin{pmatrix}
 y_1 &y_2& \cdots & y_{l-2}& y_{l-1} & G(x)
 \\
F(x) &y_1 & \cdots & y_{l-3}&y_{l-2}&y_{l-1}
\end{pmatrix}\!.
\end{gather*}
Note that the relations in~\eqref{eq:prime} give a prime ideal in $\C[x,y_1,\ldots,y_{l-1}]$.

 Inspired by the paper~\cite{KMP:13} (also see~\cite{P:74}),
we consider the following deformation,
\begin{gather}\label{eq:deformation}
F(x)\longrightarrow\tilde F(x)=\prod_{j=1}^m\big(x-\lambda_j^l\big)\qquad\text{and}\qquad
G(x)\longrightarrow \tilde G(x)=x\prod_{j=1}^m\big(x-\lambda_{m+j}^l\big),
\end{gather}
where $\lambda_{j}$ for $j=1,2,\ldots, 2m$ are all distinct.
The singular limit is then given by coalescing two moduli parameters of the space curve,
\begin{gather*}
\lambda_j^l,\,\lambda_{m+j}^l\longrightarrow \kappa_j^l\qquad \text{for}\quad j=1,\ldots,m.
\end{gather*}
One should note here that the singularity obtained by the above process is an \emph{ordinary $l$-tuple
point singularity}~\cite[p.~247]{BG:80}. When $l=2$, the hyperelliptic case, this is an ordinary double point singularity~\cite{Mu:84, SW:85}.

The smoothness of the curve can be shown as follows:
We consider a commutative ring defined by
\begin{gather*}
\mathcal{R}=\C[x,y_1,y_2,\ldots,y_{l-1}]/\mathcal{P},
\end{gather*}
where the prime ideal $\mathcal{P}$ is given by the relations in~\eqref{eq:prime},
\begin{gather*}
\mathcal{P}=\{p_{i,j}=y_iy_{j-1}-y_jy_{i-1}=0\colon 1\le i<j\le l\}\qquad
\text{with}\quad
y_0=\tilde F(x),\quad
y_l=\tilde G(x).
\end{gather*}
(Hereafter we will omit the $\,\widetilde{}\,$ on the functions $F(x)$ and $G(x)$.)
Then the affine part of the curve associated with the soliton solution is given by $\text{Spec}(\mathcal{R})$. We~now show that $\Sp(\RR)$ is non-singular.

\begin{Proposition}\label{prop:smooth}
For every $(x,y_1,\ldots,y_{l-1})$ satisfying all $p_{i,j}=0$, the following Jacobian $\mathcal{U}$ has rank $l-1$,
\begin{gather*}
\mathcal{U}:=\bigg(
\frac{\partial}{\partial x}p_{i,j},\frac{\partial}{\partial y_1}p_{i,j}, \ldots, \frac{\partial}{\partial y_{l-1}}p_{i,j}\bigg)_{1\le i<j\le l}.
\end{gather*}
Here $p_{i,j}$ in each column of the matrix $\mathcal{U}$ are arranged in the following way,
\begin{gather*}
\big(p_{1,j}~(2\le j\le l);~p_{2,j}~(3\le j\le l);~\ldots;~p_{i,j}~(i+1\le j\le l);~\ldots;~p_{l-1,l}\big)^{\rm T}.
\end{gather*}
\end{Proposition}

\begin{proof}
The $\binom{l}{2}\times l$ Jacobian matrix has the following structure.\vspace{2ex}

(1) For the columns with $p_{1,j}$ ($2\le j\le l$), we have the $(l-1)\times l$ matrix,
\begin{gather*}
\begin{pmatrix}
-y_2F' &2y_1&-F& 0&0&\cdots & 0 \\
-y_3F' &y_2&y_1&-F&0&\cdots &0\\
-y_4F' & y_3 &0 & y_1 & -F&\cdots & 0 \\
\vdots & \vdots& \vdots&\ddots &\ddots& \ddots&\vdots\\
-y_{l-1}F'&y_{l-2}&0&\cdots &0& y_1&-F\\
-(FG)'&y_{l-1}&0&\cdots& 0&0& y_1
\end{pmatrix}\!.
\end{gather*}\vspace{2ex}

(2) For the columns with $p_{i,j}$ $(i+1\le j\le l)$ and $2\le i\le l-3$, we have the $(l-i)\times l$ matrix,
\setcounter{MaxMatrixCols}{11}
\begin{gather*}
\begin{pmatrix}
0&0 & \cdots & 0& -y_{i+1} & 2 y_i & -y_{i-1}& 0 &0&\cdots &0\\
0&0&\cdots & 0 & -y_{i+2}&y_{i+1}&y_i& -y_{i-1} & 0 &\cdots &0\\
0&0&\cdots&0& -y_{i+3}&y_{i+2}& 0 &y_i&-y_{i-1}&\cdots & 0 \\
\vdots&\vdots &\vdots &\vdots &\vdots &\vdots &\vdots&\ddots&\ddots&\ddots&\vdots\\
0&0 &\cdots &0 &-y_{l-1}& y_{l-2}& 0&\cdots &0&y_i&-y_{i-1}\\
-y_{i-1}G'&0&\cdots&0&-G& y_{l-1} &0 &\cdots&\cdots &0 &y_i
\end{pmatrix}\!,
\end{gather*}
where the first nonzero entry in each row corresponds to $\partial p_{i,j}/\partial y_{i-1}$.\vspace{2ex}

(3) For the last three rows, we have
\begin{gather*}
\begin{pmatrix}
0&0&\cdots &0& -y_{l-1}& 2y_{l-2}&-y_{l-3}\\
-y_{l-3}G'&0&\cdots&0&-G&y_{l-1}&y_{l-2}\\
-y_{l-2}G'&0&\cdots&0&0&-G&2y_{l-1}
\end{pmatrix}\!.
\end{gather*}

We consider the cases: $(a)$ $F(x)=0$, $(b)$ $G(x)=0$, and $(c)$ $F(x)G(x)\ne 0$.

\begin{itemize}\itemsep=0pt
\item[$(a)$] $F(x)=0$ implies $x=\lambda_j$ for some $1\le j\le m$. Then, we have $y_i=0$ for all $i=1,\dots,l-1$.
Then the Jacobian matrix has the following nonzero entries,
\begin{gather*}
\frac{\partial p_{1,l}}{\partial x}\bigg|_{F=0,y=0}=-F'G,\qquad
\frac{\partial p_{i,l}}{\partial y_{i-1}}\bigg|_{F=0,y=0}=-G \qquad(2\le i\le l-1).
\end{gather*}
This implies ${\rm rank}(\mathcal{U})=l-1$.
\item[$(b)$] $G(x)=0$ implies $x=0$ or $x=\lambda_{m+j}$ for some $1\le j\le m$. Then, we have $y_i=0$ for all $i=1,\dots,l-1$.
Then the Jacobian matrix has the following nonzero entries,
\begin{gather*}
\frac{\partial p_{1,l}}{\partial x}\bigg|_{F=0,y=0}=-FG',\qquad
\frac{\partial p_{1,j}}{\partial y_{j}}\bigg|_{F=0,y=0}=-F\qquad
(2\le j\le l-1).
\end{gather*}
This implies ${\rm rank}(\mathcal{U})=l-1$.
\item[$(c)$] For the case $F(x)G(x)\ne 0$, we have all non zero variables $(x,y_1,\ldots,y_{l-1})$. Let us consider the null space of $\mathcal{U}^{\rm T}$, the transpose of $\mathcal{U}$.
Each vector in ${\rm null}\big(\mathcal{U}^{\rm T}\big)$ can be found is the following form: For $3\le i\le l$, we have $(l-i+1)\times \binom{l}{2}$ matrix whose rows satisfy $r\mathcal{U}=(0,\ldots,0)$,
\setcounter{MaxMatrixCols}{17}
\begin{gather*}\setlength{\arraycolsep}{4pt}
\begin{pmatrix}
0&\cdots&0&y_{i}& -y_{i-1}& 0&\cdots &0& y_1& 0&\cdots &\cdots&\cdots&\cdots &0\\
0&\cdots&0&y_{i+1}& 0& -y_{i-1}& 0&\cdots &0 &y_1&0 &\cdots&\cdots& \cdots&0\\
\vdots&\vdots&\vdots&\vdots &\vdots &\ddots &\ddots &\ddots &\vdots &\ddots& \ddots &\ddots&\cdots &\cdots&\vdots\\
0&\cdots&0&y_{l-1}&0& \cdots &0 &-y_{i-1}& 0& \cdots &0&y_1 &0& \cdots &0\\
0&\cdots&0&y_l& 0 &\cdots &0&0&-y_{i-1}& 0 &\cdots&0 &y_1&\cdots&0
\end{pmatrix}\!,
\end{gather*}
where the first nonzero entry in each row is at $(i-2)$-th place, and $y_l=G$. The total number of these rows is $\binom{l-1}{2}$. This implies that the nullity of $\mathcal{U}^{\rm T}$ is $\binom{l-1}{2}$, hence
the rank of $\mathcal{U}$ is $\binom{l}{2}-\binom{l-1}{2}=l-1$.
\end{itemize}
This proves that the affine variety given by $\Sp(\RR)$ is a smooth curve.
\end{proof}

A Riemann surface associated with the curve $\mathcal{C}$ can be obtained by adding one point $\infty$ to the affine smooth curve given by $\Sp(\RR)$. At $\infty$, we introduce
the variables,
\begin{gather*}
\underline{x}=\frac{1}{x},\qquad
\text{and}\qquad
\underline{y}_i=\frac{y_i}{x^{m+1}}\qquad
\text{for}\quad 1\le i\le l-1.
\end{gather*}
Then we have a commutative ring,
\begin{gather*}
\underline{\RR}=\C[\underline{x},\underline{y}_1,\ldots,\underline{y}_{l-1}]/\underline{\mathcal{P}},
\end{gather*}
where the prime ideal $\underline{\mathcal{P}}$ is given by
\begin{gather*}
\underline{\mathcal{P}}=\left\{\underline{p}_{i,j}
=\left|\begin{matrix}\underline{y}_i&\underline{y}_j\\
\underline{y}_{i-1}&\underline{y}_{j-1}\end{matrix}\right|\!\colon 1\le i<j\le l\right\}\!,
\end{gather*}
with
\begin{gather*}
\underline{y}_0=\underline{x}\prod_{j=1}^m\big(1-\lambda_j^l\underline{x}\big),\qquad
\underline{y}_l=\prod_{j=1}^m\big(1-\lambda_{m+j}^l\underline{x}\big).
\end{gather*}
In a similar manner, we can prove that the affine curve given by $\Sp(\underline{\RR})$ is
also smooth. Then the Riemann surface associated to the curve $\mathcal{C}$ is obtained by
patching these affine curves.
In~terms of the notion of the Weierstrass semigroup~\cite{RG:09}, we have the following corollary.
\begin{Corollary}
The numerical semigroup of type $\langle l, lm+1, lm+2,\ldots, l(m+1)-1\rangle$ is Weierstrass for $m\ge2$.
\end{Corollary}

Finally, we remark that the spectrum $\Sp(\RR)$ is given by a projection of
$\Sp\big(\widehat{\RR}\big)$ with the following ring $\widehat{\RR}$,
\begin{gather*}
\widehat{\RR}:=\C[x,w_1,w_2]/\big(w_1^l-F(x), w_2^l-G(x)\big).
\end{gather*}
We note a natural ring homomorphism $\RR\to\widehat{\RR}$ with
\begin{gather*}
y_i=w_1^{l-i}w_2^i\qquad\text{for}\quad i=1,\ldots,l-1.
\end{gather*}
This leads to a projection,
\begin{gather*}
\Sp\big(\widehat{\RR}\big)\longrightarrow \Sp(\RR).
\end{gather*}

We conclude the section with the following remark about the cases for the generalized
soliton solutions, i.e., $k<l-1$.
\begin{Remark}
In the case of $k=1$ in~\eqref{eq:DegenerateIdeal}, we have just one relation,
\begin{gather*}
\begin{pmatrix}
y_1 &G(x)\\
F(x)^{l-1} & y_1^{l-1}
\end{pmatrix}\qquad\text{which gives}\quad y_1^l=G(x)F(x)^{l-1},
\end{gather*}
which is a singular plane curve. One can deform the curve with
\begin{gather*}
\begin{cases}
F(x)^{l-1}\longrightarrow \displaystyle{\prod_{i=0}^{l-2}F_{i+1}(x)\qquad\text{with}\quad F_{i+1}}(x)=\prod_{j=1}^m\big(x-\lambda_{im+j}^l\big),
\\[2.0ex]
G(x)\longrightarrow \displaystyle{x\prod_{j=1}^m\big(x-\lambda_{lm-j}^l\big)},
\end{cases}
\end{gather*}
which then gives a smooth plane curve, called cyclic $(l,lm+1)$-curve,
\begin{gather*}
y_1^l=G(x)\prod_{i=1}^{l-1}F_i(x)=x\prod_{j=1}^{lm}\big(x-\lambda_j^l\big).
\end{gather*}
The cases with $l=2$ are just the hyperelliptic ones. The singular curve associated to the generalized soliton solution is obtained by coalescing $l$ points of the $(l,lm+1)$-curve,
\begin{gather*}
\lambda_{j}^l,\lambda_{m+j}^l,\ldots,\lambda_{(l-1)m+j}^l \longrightarrow \kappa_j^l\qquad
\text{for}\quad j=1,\ldots,m.
\end{gather*}

For the cases with $1<k<l-1$, it seems in general that
the singular curve is not smoothable in $\C\mathbb{P}^{k+1}$~\cite{G:82, P:74}. One of a special case was found in~\cite{KMP:13} for $l=6$, $m=2$ and $k=4$, i.e., the corresponding numerical semigroup is of type $\langle 6,13,14,15,16\rangle$. One can also show that the case of $\langle 4,4m+1,4m+2\rangle$ (i.e., $l=4$ and $k=2$) has the following ``smooth'' deformation. The ring of the polynomials $\RR$ is given by
$\C[x,y_1,y_2]/{\tilde{\mathcal{P}}}$, where the prime ideal is given by
\begin{gather*}
\big\{y_1^2-y_2F_2(x), y_2^2-G(x)F_1(x)\big\},\qquad
\text{from}\quad\begin{pmatrix}
y_1 &y_2 & G(x)\\
F_1(x)F_2(x)& F_1(x)y_1&y_1y_2
\end{pmatrix}\!,
\end{gather*}
where $F_1(x)$, $F_2(x)$ and $G(x)$ are given by
\begin{gather*}
F_1(x)=\prod_{j=1}^m\big(x-\lambda_j^4\big),\qquad
F_2(x)=\prod_{j=1}^m\big(x-\lambda_{m+j}^4\big),\qquad
G(x)=x\prod_{j=1}^m\big(x-\lambda_{2m+j}^4\big).
\end{gather*}
This smooth curve is an irreducible component of the intersection of the hypersurfaces given by
\begin{gather*}
y_1^4=G(x)F_1(x)F_2(x)^2\qquad\text{and}\qquad y_2^2=G(x)F_1(x).
\end{gather*}

We will discuss the deformation problem for the cases with $2\le k\le l-2$ for $l\ge 5$ in a future communication.
\end{Remark}

\section{Spectral curves for soliton solutions}\label{sec:spectralC}
In this section, we identify the singular space curve given by~\eqref{eq:prime} as a spectral curve
of commuting differential operators in the Lax--Sato formulation of the KP hierarchy (see Appendix~\ref{A}, and also, e.g.,~\cite{K:17} for the formulation). This is an extension of the well-know theorem
by~Bur\-ch\-nall and Chaundy~\cite{BC:23}. We~here consider the cases for the soliton solutions
corresponding to~$k=l-1$ and the generalized soliton solutions corresponding to $k=l-2$. The general case with $1\le k\le l-3$ will be left for the readers.

\subsection[Spectral curve for the soliton solution (k=l-1)]{Spectral curve for the soliton solution ($\boldsymbol{k=l-1}$)}

We first recall that the $\tau$-function of the soliton solution is given by $\tau(t)=\big|E(t)A^{\rm T}\big|$ in~\eqref{eq:tauE}, where the $m\times lm$ matrices $A$ is given by
\begin{gather*}
A=I_m\otimes \Omega_l^1,\qquad\text{where}\quad \Omega_l^1=\big(1,\omega_l,\omega_l^2,\ldots,\omega_l^{l-1}\big)\qquad\text{with}\quad \omega_l=\exp\bigg(\frac{2\pi {\rm i}}{l}\bigg),
\end{gather*}
and $E(t)$ whose base exponential functions~\eqref{eq:base} are given by
\begin{gather*}
[\hat{\mathsf{E}}^{(0)}]_m=(\mathsf{E}(\kappa_1),\ldots,\mathsf{E}(\kappa_m))\qquad\text{with}\quad \mathsf{E}(\kappa_j)=(E_1(t,\kappa_j),\ldots,E_l(t,\kappa_j)).
\end{gather*}
Here $E_i(t,\kappa_j)$ is the exponential function given in~\eqref{eq:Ej}, i.e.,
\begin{gather*}
E_i(t,\kappa_j)=\exp\bigg(\sum_{n=1}^\infty \big(\kappa_j\omega^{i-1}\big)^nt_n\bigg).
\end{gather*}

Now we define the following time-operator,
\begin{gather*}
T_\alpha:=\sum_{i=0}^m(-1)^i\sigma_i(\kappa)\partial_{(m-i)l+\alpha}\qquad
\text{for}\quad 1\le \alpha\le l-1,
\end{gather*}
where $\sigma_i(\kappa)$ is the elementary symmetric polynomial of degree $i$ in $\big(\kappa_1^l,\ldots,\kappa_m^l\big)$, i.e.,
\begin{gather}\label{eq:chara}
\prod_{j=1}^m(\lambda-\kappa_j^l)=\sum_{i=0}^m(-1)^i\sigma_i(\kappa)\lambda^{m-i}.
\end{gather}
Then we have the following proposition.
\begin{Proposition}\label{prop:SolTau}
The $\tau$-function generated by $(A,E(t))$ above satisfies
\begin{gather*}
T_{\alpha}\tau(t)=0\qquad \text{for}\quad 1\le \alpha\le l-1.
\end{gather*}
\end{Proposition}

\begin{proof}
Since the $\tau$-function depends only on the exponential functions $E_s(t,\kappa_j)$ for $1\le s\le l$, it is sufficient to show that each $E_s(t,\kappa_j)$ satisfies
\begin{gather*}
T_{\alpha}E_s(t,\kappa_j)=0\qquad \text{for all}\quad 1\le\alpha\le l-1.
\end{gather*}
Direct computation shows
\begin{gather*}
T_{\alpha}E_s(t,\kappa_j)=\bigg(\sum_{i=0}^m(-1)^i\sigma_i(\kappa) \big(\kappa_j\omega_l^{s-1}\big)^{(m-i)l+\alpha}\bigg)E_s(t,\kappa_j).
\end{gather*}
Since $\omega_l^l=1$, the term in the parenthesis in the equation becomes
\begin{gather*}
\bigg(\sum_{i=0}^m(-1)^i\sigma_i(\kappa)\kappa_j^{(m-i)l}\bigg)\big(\omega_l^{s-1}\kappa_j\big)^\alpha=0,
\end{gather*}
where we have used~\eqref{eq:chara} with $\lambda=\kappa_j^l$. This proves the proposition.
\end{proof}
\begin{Remark}
Note here that the equation $T_\alpha\tau(t)=0$ does not depend on the matrix~$A$.
This implies that any soliton solution of the $l$-th generalized KdV hierarchy satisfies
Proposition~\ref{prop:SolTau}. Also note that the number of the free parameters in the
matrix~$A$ is $m(l-1)$, which is the genus of the corresponding numerical semigroup, i.e., $g(S)=m(l-1)$.
\end{Remark}

Since the solutions of the $l$-th generalized KdV hierarchy can be expressed by
a single $\tau$-function, they also satisfy $T_\alpha v_i=0$, where $v_i$'s are from the $l$-th differential operator $L^l$ of~\eqref{eq:Laxl}
in Appendix~\ref{A}.

Now we define the following differential operators of order $ml+\alpha$ in $\partial$,
\begin{gather*}
L_{\alpha}:=\sum_{i=0}^m(-1)^i\sigma_i(\kappa)B_{(m-i)l+\alpha}\qquad \text{for}\quad 1\le\alpha\le l-1,
\end{gather*}
where $B_\beta$ is the differential operator of order $\beta$ defined in~\eqref{eq:Lax}, i.e., $B_{\beta}=(L^\beta)_{\ge0}$. Then we have
the following proposition.

\begin{Proposition}\label{prop:CommL}
The $l$ differential operators $\big\{L^l,L_1,\ldots,L_{l-1}\big\}$ mutually commute, i.e., \begin{itemize}\itemsep=0pt
\item[$(a)$] $\big[L_\alpha,\,L^l\big]=0$ for any $1\le\alpha\le l-1$, and
\item[$(b)$] $[L_\alpha,\,L_\beta]=0$ for any $1\le \alpha,\beta\le l-1$.
\end{itemize}
\end{Proposition}

\begin{proof}
$(a)$ Using the Lax equation~\eqref{eq:Lax}, the commutator becomes
\begin{gather*}
\big[L_\alpha,L^l\big] = \sum_{i=0}^m(-1)^i\sigma_i(\kappa)\big[B_{(m-i)l+\alpha},L^l\big] =\sum_{i=0}^m(-1)^i\sigma_i(\kappa)\partial_{(m-i)l+\alpha}\big(L^l\big)=T_\alpha \big(L^l\big).
\end{gather*}
Since $T_\alpha \tau=0$ (Proposition~\ref{prop:SolTau}), we have $T_\alpha(L^l)=0$.

$(b)$ Using the Zakharov--Shabat equations in~\eqref{eq:ZS}, the commutator becomes
\begin{gather*}
[L_\alpha,\,L_\beta] = \sum_{i=0}^m\sum_{j=0}^m(-1)^{i+j}\sigma_i(\kappa)\sigma_j(\kappa)[B_{(m-i)l+\alpha},B_{(m-j)l+\beta}]
\\ \hphantom{[L_\alpha,\,L_\beta]}
=\sum_{j=0}^m(-1)^j\sigma_j(\kappa)T_\alpha (B_{(m-j)l+\beta})
-\sum_{i=0}^m(-1)^i\sigma_i(\kappa)T_\beta (B_{(m-i)l+\alpha}).
\end{gather*}
Again, note that the coefficients of the differential operators $B_n$ depend on $t$ only through the $\tau$-function.
This implies that $T_\alpha (B_\beta)=0$ for any $1\le \alpha,\beta\le l-1$.

This completes the proof.
\end{proof}
\begin{Remark}
The statement $(b)$ in Proposition~\ref{prop:CommL} is a direct consequence of $(a)$ due to a result of Schur~\cite{Sc:05} (see also~\cite{Ml:90}).
\end{Remark}

In order to find the spectral curve given by the commuting differential operators $\big\{L^l,L_1$, $\ldots,L_{l-1}\big\}$, we first note the following lemma.
\begin{Lemma}\label{lem:Spec}
For the wave function $\phi=W\phi_0$ in~\eqref{eq:phiW}, we have
\begin{gather*}
L_\alpha\phi=\bigg(z^{-ml-\alpha}\prod_{j=1}^m\big(1-(\kappa_jz)^l\big)\bigg)\phi\qquad
\text{for}\quad 1\le \alpha\le l-1.
\end{gather*}
\end{Lemma}
\begin{proof} We first note that
\begin{gather*}
L_\alpha\phi=\sum_{i=0}^m(-1)^i\sigma_i(\kappa)B_{(m-i)l+\alpha}\phi=T_{\alpha}\phi,
\end{gather*}
where we have used $\partial_n\phi=B_n\phi$ in~\eqref{eq:LaxPair}.
Then we write the wave function in the dressing form $\phi=W\phi_0$ as in~\eqref{eq:wave} with~\eqref{eq:phiW}.
Noting again that the dressing operator depends on $t$ only through the $\tau$-function, we have
\begin{gather*}
T_\alpha\phi=WT_{\alpha}\phi_0=W\bigg(\sum_{i=0}^m(-1)^i\sigma_i(\kappa)z^{-(m-i)l-\alpha}\bigg)\phi_0
=z^{-ml-\alpha}\bigg(\sum_{i=0}^m(-1)^i\sigma_i(\kappa)z^{il}\bigg)\phi
\\ \hphantom {T_\alpha\phi}
{}=z^{-ml-\alpha}\prod_{j=1}^m\big(1-(\kappa_jz)^l\big)\phi.
\end{gather*}
This proves the lemma.
\end{proof}

Now using the coordinates in~\eqref{eq:coordinates}, i.e.,
\begin{gather*}
x=z^{-l},\qquad
y_\alpha=z^{-ml-\alpha}\prod_{j=1}^m\big(1-(\kappa_jz)^l\big)\qquad
\text{for}\quad \alpha=1,\ldots,l-1,
\end{gather*}
we have
\begin{gather*}
L^l\phi=x\phi,\qquad
L_{\alpha}=y_\alpha\phi\qquad
\text{for}\quad \alpha=1,\ldots,l-1.
\end{gather*}
Then from Lemma~\ref{lem:Spec}, the following theorem is immediate (see Section~\ref{sec:deformation}).
\begin{Theorem}\label{thm:SpecEV}
The eigenvalues of the commuting differential operators $\{L^l,L_1,\ldots,L_{l-1}\}$ satisfy
the following relations,
\begin{gather*}
p_{i,j}=y_iy_{j-1}-y_jy_{i-1}=0\qquad
\text{for}\quad 1\le i<j\le l,
\end{gather*}
where $y_0=F(x)=\prod_{j=1}^m\big(x-\kappa_j^l\big)$ and $y_l=G(x)=xF(x)$.
\end{Theorem}

\begin{Remark}
Theorem~\ref{thm:SpecEV} can be considered as an extension of the well-known theorem by~Bur\-ch\-nall and Chaundy~\cite{BC:23} on the commuting pair of differential operators of positive order.
\end{Remark}

\subsection[Spectral curve for the generalized soliton solution with k=l-2]{Spectral curve for the generalized soliton solution with $\boldsymbol{k=l-2}$}

In this case, we have $q=n_{l,k}=\big\lceil\frac{l-1}{l-2}\big\rceil=2$, hence the matrix~$A$ in~\eqref{eq:Amatrix} is an $lm\times 2lm$ matrix given by
\begin{gather*}
A=\begin{pmatrix}
I_m\otimes[\Omega_l]^1 & 0\\
0&I_m\otimes[\Omega_l]^{l-1}
\end{pmatrix}\qquad
\text{with}\quad [\Omega_l]^n=\begin{pmatrix}\Omega_l^1\\ \vdots\\\Omega_l^n\end{pmatrix}\!.
\end{gather*}
The base functions in~\eqref{eq:base} are given by
\begin{gather*}
\big(\big[\hat{\mathsf{E}}^{(1)}\big]_m,\big[\hat{\mathsf{E}}^{(0)}\big]_m\big)\qquad
\text{with}\quad
\big[\hat{\mathsf{E}}^{(1)}\big]_m=\big(\hat{\mathsf{E}}^{(1)}(\kappa_1),\ldots, \hat{\mathsf{E}}^{(1)}(\kappa_m)\big),
\end{gather*}
where
\begin{gather*}
 \hat{\mathsf{E}}^{(1)}(\kappa_j)=\big(E^{(1)}_1(t,\kappa_j),\ldots, E^{(1)}_l(t,\kappa_j)\big)\qquad
 \text{with}\quad
 E^{(1)}_i(t,\kappa_j)=\frac{\partial}{\partial \kappa_j}\kappa_jE_i(t,\kappa_j).
\end{gather*}
Since the matrix~$A$ and the base functions consist of $m$ copies of the single case with diffe\-rent~$\kappa_j$'s, we here consider the case for $m=1$. The general case for $m>1$ follows in the similar manner as the case with $m=1$. The $\tau$-function for this case has the following expansion,
\begin{gather*}
\tau(t)=\big|E(t)A^{\rm T}\big|=\sum_{i=1}^l\sum_{j=1}^l\Delta_{I_{i,\hat{j}}}(A)E_{I_{i,\hat{j}}}(t)\qquad
\text{with}\quad
I_{i,\hat{j}}=\big(i,l+1,\ldots,\widehat{l+j},\ldots,2l\big),
\end{gather*}
where $\widehat{l+j}$ is the missing column index of the matrix~$A$ and the matrix $E(t)$ is given by
\begin{gather*}
E(t)=\begin{pmatrix}
\hat{\mathsf{E}}^{(1)}(\kappa)&\mathsf{E}(\kappa)\\
\partial_1\hat{\mathsf{E}}^{(1)}(\kappa)&\partial_1\mathsf{E}(\kappa)\\
\vdots &\vdots\\
\partial_1^{l-1}\hat{\mathsf{E}}^{(1)}(\kappa)&\partial_1^{l-1}\mathsf{E}(\kappa)
\end{pmatrix}\!.
\end{gather*}
Then the exponential function $E_{I_{i,\hat{j}}}(t)$ is expressed by the $l\times l$ Wronskian determinant,
\begin{gather*}
E_{I_{i,\hat{j}}}(t)=\text{Wr}\big(E^{(1)}_i(t,\kappa),E_1(t,\kappa),\ldots, \widehat{E_j(t,\kappa)},\ldots,E_l(t,\kappa)\big),
\end{gather*}
where the column of ${E_j(t,\kappa)}$ is removed from the matrix $E(t)$.
Then we have the following proposition.
\begin{Proposition}\label{prop:GStau}
The $\tau$-function for the generalized soliton solution of type $\langle l,l+1,\ldots,2l-2\rangle$ satisfies
\begin{gather*}
T_\alpha\tau(t)=0\qquad\text{for}\quad 1\le \alpha\le l-2,
\end{gather*}
where $T_{\alpha}=\partial_{l+\alpha}-\kappa^l\partial_\alpha$.
\end{Proposition}

\begin{proof}
 First recall that $T_\alpha E_i(t,\kappa)=0$ for any $i$. Then we have
\begin{gather*}
T_\alpha E_i^{(1)}(t,\kappa)=-\frac{\partial T_\alpha}{\partial\kappa}\kappa E_i(t,\kappa)=l\kappa^{l+\alpha}\big(\omega_l^{i-1}\big)^\alpha E_i(t,\kappa).
\end{gather*}
We then note that $T_\alpha E_{I_{i,\hat{j}}}$ assumes nonzero value only if $i=j$, and
\begin{gather*}
T_\alpha E_{I_{i,\hat{i}}}(t)=(-1)^{i-1}l\kappa^{l+\alpha}\big(\omega_l^{i-1}\big)^\alpha \text{Wr}(E_1(t,\kappa),\ldots,E_l(t,\kappa)).
\end{gather*}
Now applying $T_\alpha$ on the $\tau$-function, we have
\begin{gather*}
T_\alpha\tau(t)=l\kappa^{l+\alpha}\bigg(\sum_{i=1}^l(-1)^{i-1}\big(\omega_l^{i-1}\big)^\alpha \Delta_{I_{i,\hat{i}}}(A)\bigg)\text{Wr}(E_1(t,\kappa),\ldots,E_l(t,\kappa)).
\end{gather*}
The term in the parenthesis is expressed by the $l\times l$ determinant,
\begin{gather*}
\left|\begin{matrix}
1 &\omega_l^{\alpha+1} &\omega_l^{2(\alpha+1)} &\cdots & \omega_l^{(l-1)(\alpha+1)}\\
1&\omega_l & \omega_l^2 &\cdots &\omega_l^{l-1}\\
\vdots & \vdots&\vdots&\ddots &\vdots\\
1&\omega_l^{l-1} &\omega_l^{l-2}&\cdots &\omega_l
\end{matrix}\right|
\end{gather*}
which vanishes for $1\le\alpha\le l-2$, and not zero for $\alpha=l-1$.
This completes the proof.
\end{proof}

\begin{Remark}
Unlike the case of the soliton solution, one should note that for the generalized soliton solution,
$T_\alpha\tau=0$ is true only for $1\le\alpha\le l-2$ and depends also on the matrix~$A$.
\end{Remark}

Proposition~\ref{prop:GStau} implies that the generalized soliton solutions of
type $\langle l,l+1,\ldots,2l-2\rangle$ also have
Proposition~\ref{prop:CommL} and Lemma~\ref{lem:Spec} but only for $1\le \alpha\le l-2$.

\appendix
\section{The Lax--Sato formulation of the KP hierarchy}\label{A}
Here we give a brief summary of the Lax--Sato formulation of the KP hierarchy and
the $l$-th generalized KdV hierarchy. We~also give the Wronskian
formula of the $\tau$-function for the KP solitons.

\subsection[The l-th generalized KdV hierarchy]{The $\boldsymbol l$-th generalized KdV hierarchy}

The Sato theory of the KP hierarchy is formulated on the basis of a pseudo-differential operator,
\begin{gather*}
L=\partial + u_2\partial^{-1}+u_3\partial^{-2}+\cdots,
\end{gather*}
where $\partial$ is a derivative satisfying $\partial\partial^{-1}=\partial^{-1}\partial=1$ and the generalized Leibniz rule,
\begin{gather*}
\partial^{\nu}f\cdot=\sum_{k=0}^\infty\binom{\nu}{k}\big(\partial_1^kf\big)\partial^{\nu-k}\cdot,
\end{gather*}
for any smooth functions $f$. (Note that the series terminates if and only if $\nu$ is a nonnegative integer.) Then the KP hierarchy can be written in the Lax form,
\begin{gather}\label{eq:Lax}
\partial_n(L)=[B_n,\,L]\qquad\text{with}\quad B_n=(L^n)_{\ge 0}\qquad (n=1,2,\ldots),
\end{gather}
where $(L^n)_{\ge 0}$ represents the polynomial (differential) part of $L^n$ in $\partial$.
The solution of the KP equation~\eqref{eq:KP} is given by $u=2u_2$. The Lax equation~\eqref{eq:Lax}
is also given by the compatibility condition of the linear system,
\begin{gather}\label{eq:LaxPair}
L\phi=z^{-1}\phi,\qquad
\partial_n\phi=B_n\phi,
\end{gather}
where $\phi$ is called the wave function of the Lax pair $(L,B_n)$. The compatibility among the equations $\partial_n\phi=B_n\phi$, which is $\partial_n\partial_m\phi=\partial_m\partial_n\phi$, gives
\begin{gather}\label{eq:ZS}
\partial_m(B_n)-\partial_n(B_m)+[B_n,B_m]=0,
\end{gather}
which is called the Zakharov--Shabat equations.

The variable $z\in\C$ in~\eqref{eq:LaxPair} may be considered as a local coordinate at $\infty$
in the spectral space of $L$. Note that if the functions $u_i$'s
are all zero, then we have $L=\partial$ and $B_{n}=\partial^n$ and the wave function, denoted by $\phi_0$, is given by
\begin{gather}\label{eq:phiW}
\phi_0(z;t)=\exp\bigg(\sum_{n=1}^\infty\frac{t_n}{z^n}\bigg).
\end{gather}
The wave function $\phi$ is then expressed in the dressing form,
\begin{gather}\label{eq:wave}
\phi=W\phi_0\qquad\text{with}\quad W=1-w_1\partial^{-1}-w_2\partial^{-2}-\cdots,
\end{gather}
where the pseudo-differential operator $W$ is called the dressing operator. Notice that all the functions $u_i$'s in $L$ can be determined by $w_j$'s in $W$ through
\begin{gather*}
L=W\partial W^{-1}.
\end{gather*}
For example, we have
\begin{gather*}
u_2=\partial_1w_1,\qquad
u_3=\partial_1w_2+w_1\partial_1w_1,\qquad
\ldots.
\end{gather*}
Then, from the Lax equation, the dressing operator $W$ satisfies
\begin{gather}\label{eq:Sato}
\partial_n(W)=B_nW-W\partial^n\qquad
\text{for}\quad n=1,2,\dots,
\end{gather}
which is sometimes called the Sato equation.

The $l$-th generalized KdV hierarchy is the $l$-reduction of the KP hierarchy defined by
\begin{gather*}
L^l=\big(L^l\big)_{\ge 0},
\end{gather*}
that is, the $l$-th power of $L$ becomes a differential operator. This means that the functions $u_i$'s
are determined by $l-1$ variables in $L^l$ in the form,
\begin{gather}\label{eq:Laxl}
L^l=\partial^l+v_2\partial^{l-2}+v_3\partial^{l-3}+\cdots+v_{l-1}\partial+v_l.
\end{gather}
Also note that those variables are determined by $w_i$'s in $W$. From~\eqref{eq:Lax}, the $l$-reduction gives the constraints,
\begin{gather*}
\partial_{nl}(L)=0\qquad\text{for}\quad n=1,2,\ldots,
\end{gather*}
that is, all the variables $v_i$'s do not depend on the times $t_{nl}$. The original KdV hierarchy
is given by the 2-reduction, and the solutions do not depend on the times $t_{2n}$.

\subsection[N truncation and the tau-function]{$\boldsymbol N$ truncation and the $\boldsymbol \tau$-function}

Here we explain the Wronskian formula of the $\tau$-function.
First note that a finite truncation of~$W$ with some positive integer $N$, given by
\begin{gather*}
W=1-w_1\partial^{-1}-w_2\partial^{-2}-\cdots-w_N\partial^{-N} ,
\end{gather*}
is invariant under~\eqref{eq:Sato}. We~then consider the $N$-th order differential equation,
\begin{gather*}
W\partial^N f=f^{(N)}-w_1f^{(N-1)}-w_2f^{(N-2)}-\cdots-w_Nf=0,
\end{gather*}
where $f^{(n)}=\partial_1^n f$. Let~$\{f_i\colon i=1,\ldots,N\}$ be a fundamental set of solutions of
the equation $W\partial^Nf=0$. Then the functions $w_i$'s are given by
\begin{gather*}
w_i=-\frac{1}{\tau}p_i\big({-}\tilde\partial\big)\tau\qquad
\text{for}\quad i=1,\ldots,N,
\end{gather*}
where $p_i(x)$ is the elementary Schur polynomial of degree $i$ and $\tilde\partial=\big(\partial_1,\frac{1}{2}\partial_2,\frac{1}{3}\partial_3,\ldots\big)$. Here $\tau$ is the $\tau$-function given by
the Wronskian form,
\begin{gather}\label{eq:WrA}
\tau=\text{Wr}(f_1,f_2,\ldots,f_N)=\left|\begin{matrix}
f_1 & f_2 & \cdots & f_N \\
\partial_1f_1&\partial_1f_2&\cdots&\partial_1f_N\\
\vdots &\vdots &\ddots &\vdots\\
\partial_1^{N-1}f_1&\partial_1^{N-1}f_2&\cdots&\partial_1^{N-1}f_N
\end{matrix}\right|.
\end{gather}
For the time-evolution of the functions $f_i$, we consider the following (diffusion) hierarchy,
\begin{gather*}
\partial_nf_i=\partial_1^nf_i\qquad \text{for}\quad 1\le i\le N,\quad n\in\N,
\end{gather*}
which gives the solution of the Sato equation~\eqref{eq:Sato}. Then the solution of the KP equation
can be expressed in terms of the $\tau$-function by
\begin{gather*}
u(t)=2u_2(t)=2\partial_1w_1(t)=2\partial_1^2\ln\tau(t).
\end{gather*}

\subsection*{Acknowledgements}

One of the authors (YK) would like to thank Jing-Ping Wang and Sasha Mikhailov for useful discussions at the beginning stage of the present work. He also appreciated their financial support by the EPSRC grants EP/P012698/1 and EP/P1012655/1 during his stay at University of Kent and University of Leeds. He also thanks Atsushi Nakayashiki for many helpful and useful comments on the Sato Grassmannians and his interest in the present work, and Shigeki Matsutani for valuable comments on space curves and numerical semigroups. We~would like to thank Herb Clemens for useful discussion on the singularity of space curves. We~would also like to thank the referees for valuable comments and suggestions. The present work is partially supported by NSF grant DMS-1714770.

\pdfbookmark[1]{References}{ref}
\LastPageEnding

\end{document}